\documentclass[12pt]{article}
\usepackage{a4wide,epsfig,amsmath,color}

\usepackage{axodraw}
\usepackage{feynarts}
\usepackage{epsfig}
\usepackage{hyperref}
\usepackage[all]{hypcap}
\usepackage{amsmath}

\newcommand{\agt}{\rlap{\lower 3.5 pt \hbox{$\mathchar \sim$}} \raise 1pt
 \hbox {$>$}}
\newcommand{\alt}{\rlap{\lower 3.5 pt \hbox{$\mathchar \sim$}} \raise 1pt
 \hbox {$<$}}

\newcommand{\tr}{\mathop{\mbox{tr}}\nolimits}

\catcode`@=11
\newcount\@tempcntc
\def\@citex[#1]#2{\if@filesw\immediate\write\@auxout{\string\citation{#2}}\fi
  \@tempcnta\z@\@tempcntb\m@ne\def\@citea{}\@cite{\@for\@citeb:=#2\do
    {\@ifundefined
       {b@\@citeb}{\@citeo\@tempcntb\m@ne\@citea\def\@citea{,}{\bf
?}\@warning
       {Citation `\@citeb' on page \thepage \space undefined}}%
    {\setbox\z@\hbox{\global\@tempcntc0\csname b@\@citeb\endcsname\relax}%
     \ifnum\@tempcntc=\z@ \@citeo\@tempcntb\m@ne
       \@citea\def\@citea{,}\hbox{\csname b@\@citeb\endcsname}%
     \else
      \advance\@tempcntb\@ne
      \ifnum\@tempcntb=\@tempcntc
      \else\advance\@tempcntb\m@ne\@citeo
      \@tempcnta\@tempcntc\@tempcntb\@tempcntc\fi\fi}}\@citeo}{#1}}
\def\@citeo{\ifnum\@tempcnta>\@tempcntb\else\@citea\def\@citea{,}%
  \ifnum\@tempcnta=\@tempcntb\the\@tempcnta\else
   {\advance\@tempcnta\@ne\ifnum\@tempcnta=\@tempcntb \else
\def\@citea{--}\fi
    \advance\@tempcnta\m@ne\the\@tempcnta\@citea\the\@tempcntb}\fi\fi}
\catcode`@=12
\allowdisplaybreaks

\begin{document}

\title{
\vskip-3cm{\baselineskip14pt
\centerline{\normalsize DESY 15--063\hfill ISSN 0418-9833}
\centerline{\normalsize MPP--2015--94\hfill}
\centerline{\normalsize April 2015\hfill}}
\vskip1.5cm
Electroweak corrections to $Z$-boson hadroproduction at finite transverse
momentum}

\author{W.~Hollik,${}^a$ B.~A. Kniehl,${}^b$ E.~S. Scherbakova,${}^b$
O.~L. Veretin${}^b$\\
{\normalsize $a$ Max-Planck-Institut f\"ur Physik
(Werner-Heisenberg-Institut),}\\
{\normalsize F\"ohringer Ring 6, 80805 M\"unchen, Germany}\\
{\normalsize $b$ II. Institut f\"ur Theoretische Physik, Universit\"at
Hamburg,}\\
{\normalsize Luruper Chaussee 149, 22761 Hamburg, Germany}
}

\date{}

\maketitle

\begin{abstract}
We calculate the full one-loop electroweak radiative corrections, of
${\cal O}(\alpha^2\alpha_s)$, to the cross section of single $Z$-boson
inclusive hadroproduction at finite transverse momentum ($p_T$).
This includes the ${\cal O}(\alpha)$ corrections to $Z+j$ production, the
${\cal O}(\alpha_s)$ corrections to $Z+\gamma$ production, and certain
QCD-electroweak interference contributions involving a single quark trace.
We recover the QCD and purely weak corrections and study the QED corrections
and the QCD-electroweak interference contributions for the first time.
We also consider direct and resolved photoproduction in elastic and
inelastic scattering.
We present $p_T$ and rapidity distributions for the experimental conditions at
the Fermilab Tevatron and the CERN LHC
and assess the significance of the various contributions.
%estimate the theoretical uncertainties,
%and compare the electromagnetic and purely weak corrections
%with the QCD ones at next-to-leading order.

\medskip

\end{abstract}

\newpage

\section{Introduction}
\label{sec:one}

The study of single electroweak-gauge-boson hadroproduction, via the so-called
Drell--Yan process, has a long history, starting from the discovery of the $W$
\cite{Arnison:1983rp} and $Z$ \cite{Arnison:1983mk} bosons at the CERN Super
Proton Synchrotron (SPS) more than three decades ago, which marked a
breakthrough for the Standard Model (SM).
These processes remain to be of paramount importance also at modern hadron
colliders, such as the Fermilab Tevatron and the CERN LHC.
On the one hand, they have large cross sections and clean decay signatures in
the detectors.
This renders them particularly useful for calibrating and monitoring the
luminosities at hadron colliders, which affects all other cross section
measurements performed there as well. 
By the same token, they are sensitive probes of the parton density functions
(PDFs), in partiular of those of the quarks and antiquarks.
On the other hand, singly produced $W$ and $Z$ bosons form important
backgrounds for searches of new physics beyond the SM, such as anomalous
couplings, extra vector bosons, etc.

To achieve an adequate theoretical description, radiative corrections, both of
QCD and electroweak type, must be taken into account.
As for the total cross sections of single $W$- and $Z$-boson hadroproduction,
the next-to-leading-order (NLO) \cite{Altarelli:1978id} and
next-to-next-to-leading-order (NNLO) \cite{Matsuura:1988sm} QCD corrections
were calculated a long time ago, and also partial results at
next-to-next-to-next-to-leading-order are available \cite{Moch:2005ky}.
These corrections are of relative orders $\mathcal{O}(\alpha_s^n)$ with
$n=1,2,3$, respectively, in the strong-coupling constant $\alpha_s$.
The one-loop electroweak corrections, of relative order $\mathcal{O}(\alpha)$
in Sommerfeld's fine-structure constant $\alpha$, were studied for the $W$
boson in Ref.~\cite{Baur:1998kt} and for the $Z$ boson in
Ref.~\cite{Baur:1997wa}.
At the mixed two-loop order $\mathcal{O}(\alpha\alpha_s)$, the corrections to
the $q\bar qZ$ form factor were evaluated for light quarks $q\ne b,t$ in
Ref.~\cite{Kotikov:2007vr} using the techniques developed in
Ref.~\cite{Fleischer:1998nb}, and the treatment of the nonfactorizable
corrections in the resonance region was discussed in
Ref.~\cite{Dittmaier:2014qza}.

In order for the $W$ and $Z$ bosons to acquire finite transverse momenta
($p_T$), they must be produced in association with additional particles or
hadron jets ($j$).
The QCD corrections to the $p_T$ distributions of $W$- and $Z$-boson inclusive
hadroproduction were computed at NLO \cite{Ellis:1981hk,Gonsalves:1989ar} and
partly at NNLO \cite{Gonsalves:2005ng}.
The $\mathcal{O}(\alpha)$ corrections were investigated for the $W$ boson in
Refs.~\cite{Kuhn:2007qc,Hollik:2007sq} and for the $Z$ boson in
Refs.~\cite{Kuhn:2004em,Kuhn:2005az}.
Specifically, in Ref.~\cite{Kuhn:2007qc}, the electroweak $\mathcal{O}(\alpha)$
corrections to the $\mathcal{O}(\alpha\alpha_s)$ partonic subprocesses of
$W$-boson production were calculated imposing a minimum-transverse-momentum cut
on outgoing gluons to prevent soft-gluon singularities.
However, this cut was not applied to outgoing quarks and antiquarks as well,
which renders it impractical at the hadron level, where gluon and light-quark
jets are hard to distinguish on an event-by-event basis. 
Such a cut is also problematic from the conceptual point of view because, as
a matter of principle, a collinear gluon-photon system cannot be distinguished
from a single gluon with the same momentum.
In Ref.~\cite{Hollik:2007sq}, the results of Ref.~\cite{Kuhn:2007qc} were
confirmed, but the soft-gluon singularities were properly eliminated by
including also the $\mathcal{O}(\alpha_s)$ corrections to the
$\mathcal{O}(\alpha^2)$ partonic subprocesses.
Furthermore, the $\mathcal{O}(\alpha^3)$ contributions due to direct and
resolved photoproduction by elastic and inelastic scattering off the incoming
(anti)proton were taken into account in Ref.~\cite{Hollik:2007sq}.

In contrast to the charged-current case, the separation of the electroweak
$\mathcal{O}(\alpha)$ corrections to the neutral-current Drell--Yan process
into an electromagnetic and a weak part is meaningful with regard to infrared
(IR) and ultraviolet (UV) finiteness and gauge independence.
In Refs.~\cite{Kuhn:2004em,Kuhn:2005az}, the purely weak $\mathcal{O}(\alpha)$
corrections to the $\mathcal{O}(\alpha\alpha_s)$ partonic subprocess
$q\bar{q}\to Zg$ and its crossed versions were computed.
They make up an important subset of the contributions of absolute order
$\mathcal{O}(\alpha^2\alpha_s)$ to the inclusive hadroproduction of
finite-$p_T$ $Z$ bosons.
It is the purpose of this work to complete our knowledge of these
contributions, which have several sources, and to check the results presented
in Ref.~\cite{Kuhn:2005az}.
To start with, we need to complement the purely weak $\mathcal{O}(\alpha)$
corrections to the $\mathcal{O}(\alpha\alpha_s)$ partonic subprocess
$q\bar{q}\to Zg$ and its crossed versions by the QED ones, which have virtual
and real parts.
The $\mathcal{O}(\alpha^2\alpha_s)$ partonic subprocesses that we are then led
to consider include $q\bar{q}\to Zg\gamma$.
As in the charged-current case \cite{Hollik:2007sq}, we thus inevitably
encounter a soft-gluon singularity.
To cancel it, we need to also include the $\mathcal{O}(\alpha_s)$ QCD
corrections to the $\mathcal{O}(\alpha^2)$ partonic subprocess
$q\bar{q}\to Z\gamma$.
Furthermore, $\mathcal{O}(\alpha^2\alpha_s)$ contributions may also arise from
interferences of $\mathcal{O}(\alpha^{1/2}\alpha_s)$ and
$\mathcal{O}(\alpha^{3/2})$ Feynman diagrams yielding a single Dirac spinor
trace with nonvanishing color factor $\tr T^aT^a=N_cC_F$.
This happens for the partonic subprocess $q\bar q\to Zq\bar q$ when a diagram
involving a virtual gluon in the $s$ ($t$) channel is connected with a diagram
involving a photon or a $Z$ boson in the $t$ ($s$) channel, and for the
subprocesses $qq\to Zqq$ and $\bar q\bar q\to Z\bar q\bar q$ when a
gluon-exchange diagram and a photon/$Z$-boson-exchange diagram are connected
with twisted quark lines.
For completeness, we also recalculate the $\mathcal{O}(\alpha_s)$ QCD
corrections to the inclusive hadroproduction of finite-$p_T$ $Z$ bosons
\cite{Ellis:1981hk,Gonsalves:1989ar} and thus recover the analytic results
specified in Ref.~\cite{Gonsalves:1989ar} apart from a few misprints that we
correct.
Finally, we also study the leading-order (LO) photon-induced subprocesses, of
order $\mathcal{O}(\alpha^2)$, which, after convolution with the photon PDFs of
the (anti)proton, yield contributions of absolute order
$\mathcal{O}(\alpha^3)$. 
As in Refs.~\cite{Gonsalves:1989ar,Kuhn:2005az}, we list full analytic results
in a compact form.

Our goal is to study the inclusive hadroproduction of single $Z$ bosons with
finite values of $p_T$.
For the experimental analysis, this implies that all events with at least one
identified $Z$ boson are selected and sampled in bins of one or more kinematic
variables exclusively pertaining to the $Z$ boson, such as $p_T$ and rapidity
$y$.
If there is more than one identified $Z$ boson in such an event, then each of
them generates one entry in the considered histogram.
There is no need to identify particles of other species or jets that are
produced in association with the $Z$ bosons.
If such additional experimental information is available, it is nevertheless
ignored.
Samples of events with at least one identified $Z$ boson may, of course, also
be analyzed more exclusively.
For instance, one may study the production of a large-$p_T$ $Z$ boson in
association with a jet or a prompt photon.
The separation of $Z+j$ and $Z+\gamma$ events is efficiently achieved by means
of the procedures elaborated in studies of the photon fragmentation function
\cite{Glover:1993xc} and of photon isolation \cite{Frixione:1998jh}.
The contributions from $Z+X$ final states, in which the system $X$ contains a
heavy particle, e.g.\ a $W$, $Z$, or Higgs boson or a top quark, are greatly
suppressed and not considered here. 

The $\mathcal{O}(\alpha)$ corrections to $l^+\nu+j$, $l^+l^-+j$, and
$\nu\bar{\nu}+j$ inclusive hadroproduction were considered in
Refs.~\cite{Denner:2009gj,Denner:2011vu,Denner:2012ts}, respectively.
The theoretical study of such final states is closer to the experimental
situation, as it does not rely on the identification of the $W$ and $Z$ bosons
and the reconstruction of their four-momenta.
The latter two procedures have been routinely applied in experimental data
analyses ever since the discovery of the $W$ and $Z$ bosons at the SPS in 1983.
They are, of course, subject to certain experimental errors, which are,
however, quite small for the gold-plated $Z\to e^+e^-$ and $Z\to\mu^+\mu^-$
decay modes of relevance here. 
By the same token, the numerical results presented in Ref.~\cite{Denner:2011vu}
do not allow one to extract the $p_T$ distribution of the $Z$ boson and thus
cannot be usefully compared with the results obtained in
Refs.~\cite{Kuhn:2004em,Kuhn:2005az} and here.
This would require kinematic cuts to reduce the contributions due to the
nonresonant parts of the scattering amplitudes in Ref.~\cite{Denner:2011vu} in
analogy to the experimental acceptance cuts, e.g.\ the one confining the
invariant mass $M_{ll}$ of the $l^+l^-$ pair to an appropriately narrow
interval about $M_Z$.
Apart from that, in Ref.~\cite{Denner:2011vu}, the cross sections were not
presented as distributions in the transverse momentum of the $l^+l^-$ pair,
which could be identified with the kinematic variable $p_T$ of the $Z$ boson
for the sake of a comparison with the results obtained in
Refs.~\cite{Kuhn:2004em,Kuhn:2005az} and here.
On the other hand, the $M_{ll}$ distributions shown in Figs.~7 and 8 of
Ref.~\cite{Denner:2011vu} exhibit a rapid fall-off at the shoulders of the peak
at $M_{ll}=M_Z$ indicating that the narrow-width approximation adopted in
Refs.~\cite{Kuhn:2004em,Kuhn:2005az} and here is quite appropriate for the
Tevatron and the LHC.
The $\mathcal{O}(\alpha^2\alpha_s)$ interference contributions mentioned above
were neglected in Refs.~\cite{Denner:2011vu,Denner:2012ts} appealing to the
observation that similar contributions were found to be numerically small in
Ref.~\cite{Denner:2009gj}.
Recently, $\mathcal{O}(\alpha)$ corrections were also calculated for the
hadroproduction of the final states $l^+l^-+2j$ \cite{Denner:2014ina} and
$W^++nj$ with $n=1,2,3$ \cite{Kallweit:2014xda}.

This paper is organized as follows.
In Sec.~\ref{sec:two}, we explain our analytic calculations.
In Sec.~\ref{sec:three}, we present our numerical results.
Our conclusions are contained in Sec.~\ref{sec:four}.
Our analytic results are listed in Appendices~\ref{sec:a}--\ref{sec:c}.

%!!!
%%%%%%%%%%%%%%%%%%%%%%%%%%%%%%%%%%%%%%%%%%%%%%%%%%%%%%%%%%%%
\section{Analytic results}
\label{sec:two}

We consider the inclusive production of a $Z$ boson in the collision of two
hadrons $h_1$ and $h_2$,
\begin{equation}
h_1(P_1)+h_2(P_2)\to Z(q)+X,
\label{eq:hadron}
\end{equation}
where the four-momenta are indicated in parentheses and $X$ collectively
denotes the residual particles in the final state.
We take the $Z$ boson to be on mass shell, $q^2=M_Z^2$, neglect the hadron
masses, $P_1^2=P_2^2=0$, and define the hadronic Mandelstam variables as
\begin{equation}
  S = (P_1 +P_2)^2, \qquad
  T = (P_1-q)^2, \qquad
  U = (P_2-q)^2, \qquad
  Q^2 = q^2.
\label{eq:Mandelstam}
\end{equation}
In the center-of-mass frame, we write $q^\mu=(q^0,{\bf q}_T,q^3)$, where
${\bf q}_T$ is the transverse momentum, and define $q_T=|{\bf q}_T|$ and the
rapidity $y=(1/2)\ln[(q^0+q^3)/(q^0-q^3)]$.
Using Eq.~(\ref{eq:Mandelstam}), we have
\begin{equation}
q_T^2 = \frac{TU-Q^2(S+T+U-Q^2)}{S},\qquad
y = \frac{1}{2}\ln\frac{U-Q^2}{T-Q^2}.
\end{equation}

We work in the collinear parton model of QCD with $n_f=5$ massless quark
flavors.
We write the partonic subprocesses that contribute to the hadronic reaction in
Eq.~(\ref{eq:hadron}) generically as
\begin{equation}
i(p_1) + j(p_2) \to Z(q) + X(p_X),
\end{equation}
where $p_i=x_iP_i$ with $i=1,2$.
The partonic Mandelstam variables are defined as
\begin{equation}
s = (p_1+p_2)^2, \qquad
t = (p_1-q)^2, \qquad
u = (p_2-q)^2, \qquad
Q^2 = q^2, \qquad
s_2 = p_X^2,
\label{eq:mandelstam}
\end{equation}
and satisfy
\begin{equation}
s + t + u = Q^2 + s_2.
\end{equation}
The hadronic and partonic Mandelstam variables are related as follows: 
\begin{eqnarray}
s &=& x_1 x_2 S, \qquad t=x_1(T-Q^2)+Q^2, \qquad u=x_2 (U-Q^2)+Q^2,\nonumber\\
s_2 &=& x_1 x_2 S + x_1 (T-Q^2) + x_2 (U-Q^2) + Q^2. 
\end{eqnarray}

The differential cross section for reaction~(\ref{eq:hadron}) may be
evaluated according to
\begin{equation}
  \frac{d\sigma}{dq_T^2\,dy} = \sum_{ij} \int dx_1\,dx_2\,
     f_{i/h_1}(x_1,\mu_F^2) f_{j/h_2}(x_2,\mu_F^2)
     \frac{s\,d\sigma_{ij}}{dt\,du}(x_1 P_1, x_2 P_2, \mu_F^2),
\label{eq:xs}
\end{equation}
where the sum runs over all the partons $i$ and $j$; $f_{i/h}(x,\mu_F^2)$ is
the PDF of parton $i$ in hadron $h$; and $\mu_F$ is the factorization scale.
The partonic cross sections $d\sigma_{ij}/(dt\,du)$ may be computed in 
perturbation theory as double series in $\alpha_s$ and $\alpha$.
Apart from the Feynman rules of QCD, we also need those for the couplings of a
quark $q$ to the photon $\gamma$ and the $Z$ boson.
They are given by the vertices $ieQ_q\gamma_\mu$ and
$ie\gamma_\mu(v_q-\gamma_5 a_q)$, respectively, where $Q_q$ is the electric
charge of $q$ in units of the positron charge $e=\sqrt{4\pi\alpha}$,
\begin{equation}
v_q = \frac{I_3-2 Q_q \sin^2\theta_w}{2 \sin \theta_w\cos\theta_w}, \qquad
a_q = \frac{I_3}{2 \sin\theta_w\cos\theta_w}
\label{eq:Zqq}
\end{equation}
are its vector and axial vector couplings to the $Z$ boson, $I_3$ is its third
component of weak isospin, and $\theta_w$ is the weak mixing angle.

\begin{figure}[h]
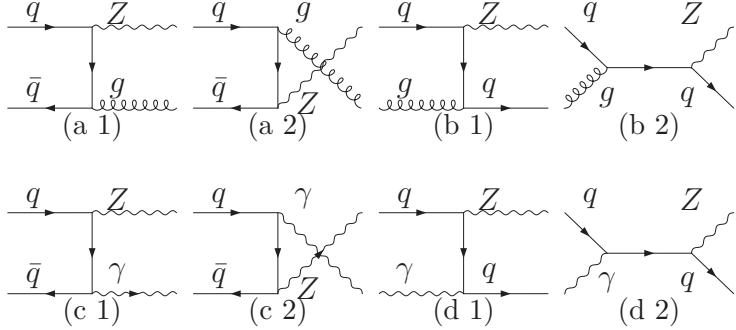

  \begin{center}
\unitlength=1.bp%
\begin{feynartspicture}(280,140)(4,2)
\input  diagBorn.inc
\end{feynartspicture}  
      \caption[]{\label{fig:born}%
Tree-level diagrams contributing to the partonic subprocesses $q+\bar q\to Z+g$
and $q+g\to Z+q$ at $\mathcal{O}(\alpha\alpha_s)$ and to the partonic
subprocesses $q+\bar q\to Z+\gamma$ and $q+\gamma\to Z+q$ at
$\mathcal{O}(\alpha^2)$.}
\end{center}
\end{figure}

In the remainder of this section, we list the relevant partonic subprocesses
with the contributing Feynman diagrams and outline the computation and its
organization.
At LO, we consider the $2\to2$ subprocesses
\begin{eqnarray}
q+\bar q&\to&Z+g,\label{qQ_Zg}\\
q+\bar q&\to&Z+\gamma,\label{qQ_Zy}\\
q+g&\to&Z+q,\label{qg_Zq}\\
q+\gamma&\to&Z+q,\label{qy_Zq}
\end{eqnarray}
where it is understood that $q$ may also be an antiquark, in which case
$\bar q$ is a quark.
They are mediated by the Feynman diagrams depicted in Fig.~\ref{fig:born}.
At LO, subprocesses (\ref{qQ_Zg}) and (\ref{qg_Zq}) are of
$\mathcal{O}(\alpha\alpha_s)$, and suprocesses (\ref{qQ_Zy}) and (\ref{qy_Zq})
are of $\mathcal{O}(\alpha^2)$.

\begin{figure}[h]
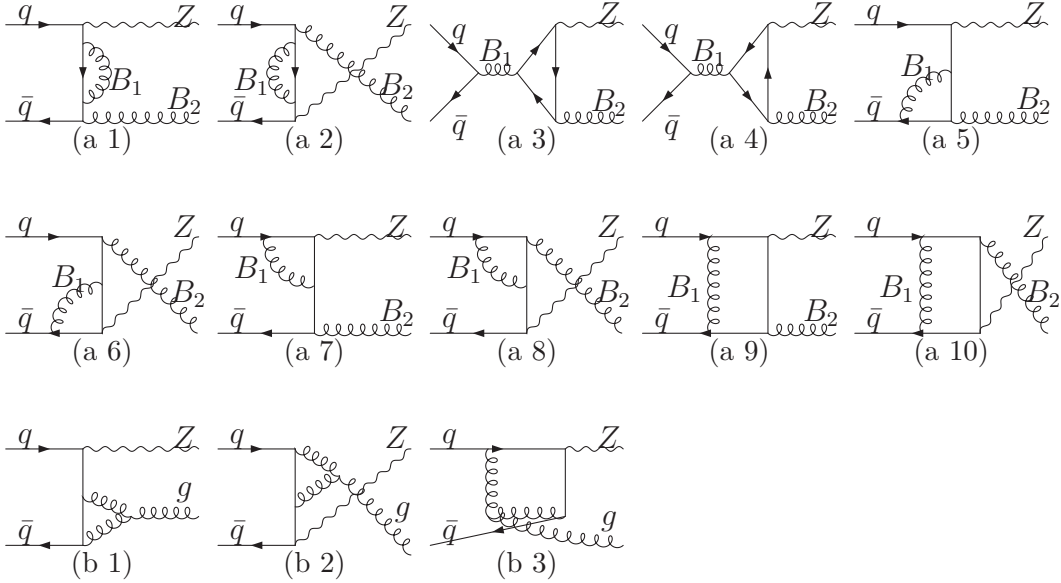

  \begin{center}
\unitlength=1.bp%
\begin{feynartspicture}(400,240)(5,3)
\input  diagLoopuU.inc
\end{feynartspicture}  
     \caption[]{\label{fig:qcd}%
QCD and QED one-loop diagrams contributing to the partonic subprocess
$q+\bar q\to Z+g$ at $\mathcal{O}(\alpha\alpha_s^2)$ ($B_1=B_2=g$) and 
$\mathcal{O}(\alpha^2\alpha_s)$ (diagrams (a~1)--(a~10) with $B_1=\gamma$ and
$B_2=g$) and to the partonic subprocess $q+\bar q\to Z+\gamma$ at
$\mathcal{O}(\alpha^2\alpha_s)$ (diagrams (a~1)--(a~10) with $B_1=g$ and
$B_2=\gamma$).
The QCD and QED one-loop diagrams contributing to the partonic subprocess
$q+g\to Z+q$ at $\mathcal{O}(\alpha\alpha_s^2)$ and
$\mathcal{O}(\alpha^2\alpha_s)$ are obtained by appropriately crossing external
legs.}
  \end{center}
\end{figure}

\begin{figure}[h]
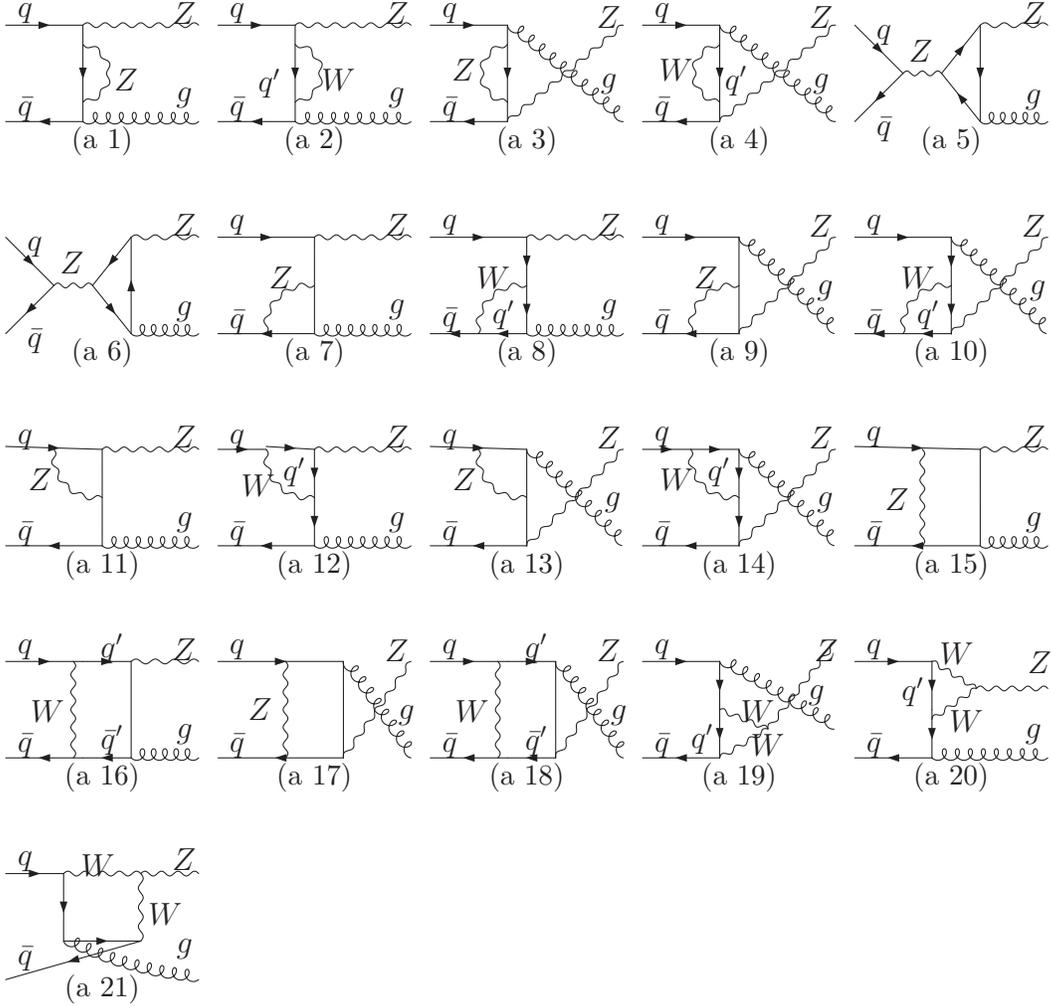

  \begin{center}
\unitlength=1.bp%
\begin{feynartspicture}(400,400)(5,5)
\input  diagLoopDdweak.inc
\end{feynartspicture}  
      \caption[]{\label{fig:weak}%
Weak-interaction one-loop diagrams contributing to the partonic subprocess
$q+\bar q\to Z+g$ at $\mathcal{O}(\alpha^2\alpha_s)$.
The weak-interaction one-loop diagrams contributing to the partonic subprocess
$q+g\to Z+q$ at $\mathcal{O}(\alpha^2\alpha_s)$ are obtained by appropriately
crossing external legs.}
  \end{center}
\end{figure}

We include the $\mathcal{O}(\alpha_s)$ QCD corrections to subprocesses
(\ref{qQ_Zg})--(\ref{qg_Zq}) and the $\mathcal{O}(\alpha)$ electroweak
corrections to subprocesses (\ref{qQ_Zg}) and (\ref{qg_Zq}), while we treat
subprocess (\ref{qy_Zq}) only at LO because of the additional
$\mathcal{O}(\alpha)$ suppression due to the photon emission by the incoming
hadrons.
The virtual QCD corrections to subprocesses (\ref{qQ_Zg}) and (\ref{qQ_Zy}) and
the virtual QED corrections to subprocess (\ref{qQ_Zg}) arise from the Feynman
diagrams in Fig.~\ref{fig:qcd}.
The virtual QCD and QED corrections to subprocess (\ref{qg_Zq}) are obtained by
appropriately crossing external legs.
The virtual weak corrections to subprocess (\ref{qQ_Zg}) arise from the Feynman
diagrams in Fig.~\ref{fig:weak}, and those to subprocess (\ref{qg_Zq}) emerge
by crossing.

\begin{figure}[h]
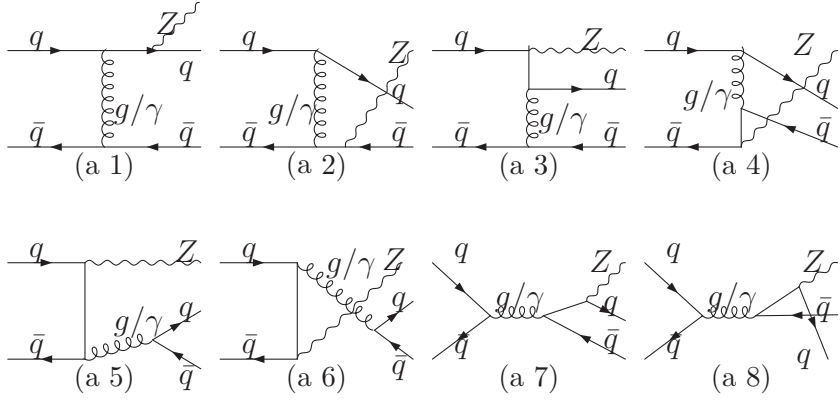

  \begin{center}
\unitlength=1.bp%
\begin{feynartspicture}(320,160)(4,2)
\input  diagBremuU.inc
\end{feynartspicture}  
      \caption[]{\label{fig:qQZqQ}%
Tree-level diagrams contributing to the partonic subprocess
$q+\bar q\to Z+q+\bar q$ at $\mathcal{O}(\alpha\alpha_s^2)$ (with a virtual
gluon).
Interferences of diagrams (a~5)--(a~8) with a virtual photon or $Z$ boson
(gluon) with diagrams (a~1)--(a~4) with a virtual gluon (photon or $Z$ boson)
contribute at $\mathcal{O}(\alpha^2\alpha_s)$.}
  \end{center}
\end{figure}

\begin{figure}[h]
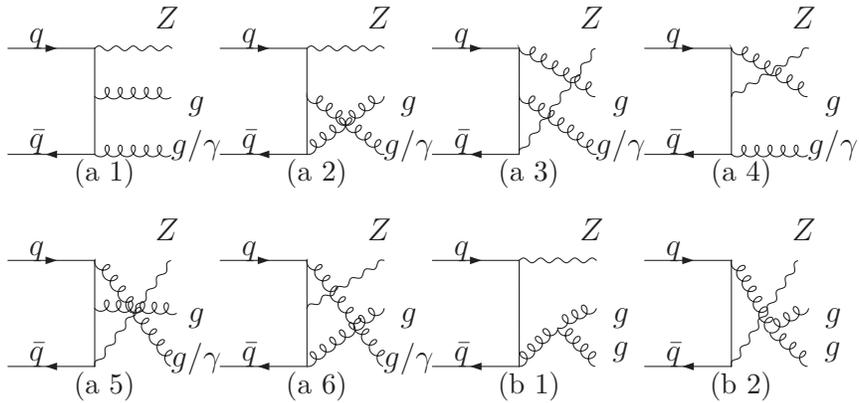

  \begin{center}
\unitlength=1.bp%
\begin{feynartspicture}(320,160)(4,2)
\input  diagBremuUgg.inc
\end{feynartspicture}  
      \caption[]{\label{fig:qQZgg}%
Tree-level diagrams contributing to the partonic subprocess $q+\bar q\to Z+g+g$
at $\mathcal{O}(\alpha\alpha_s^2)$ and to the partonic subprocess
$q+\bar q\to Z+g+\gamma$ at $\mathcal{O}(\alpha^2\alpha_s)$.}
  \end{center}
\end{figure}

\begin{figure}[h]
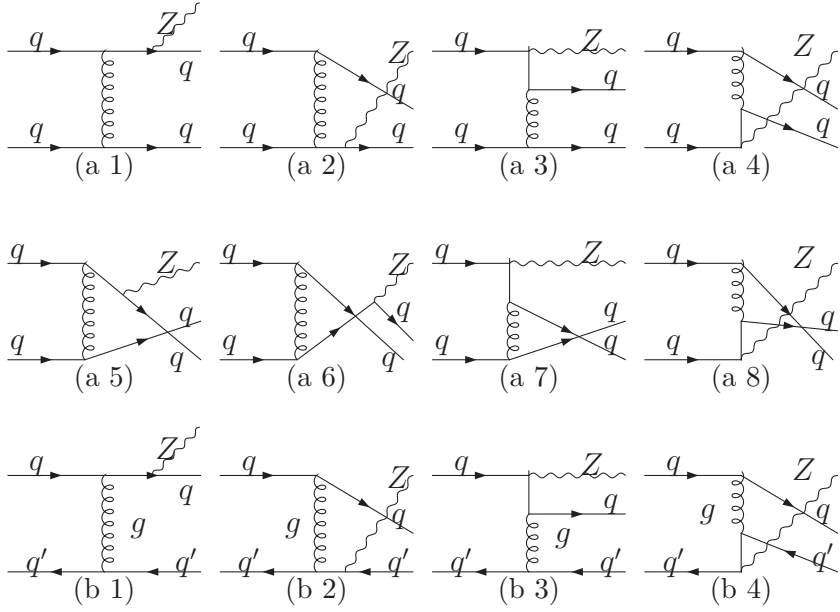

  \begin{center}
\unitlength=1.bp%
\begin{feynartspicture}(320,240)(4,3)
\input  diagBremuu.inc
\input  diagBremud.inc
\end{feynartspicture}  
      \caption[]{\label{fig:qqZqq}%
Tree-level diagrams contributing to the partonic subprocesses $q+q\to Z+q+q$
and $q+q^\prime\to Z+q+q^\prime$ at $\mathcal{O}(\alpha\alpha_s^2)$.
Interferences of diagrams (a~5)--(a~8) with the gluon replaced by a photon
or $Z$ boson with diagrams (a~1)--(a~4) contribute at
$\mathcal{O}(\alpha^2\alpha_s)$.}
  \end{center}
\end{figure}

The real QCD and QED corrections arise from the $2\to3$ subprocesses
\begin{eqnarray}
q+\bar q&\to&Z+q+\bar q,\label{qQ_ZqQ}\\
q+\bar q&\to&Z+q^\prime+\bar q^\prime,\label{qQ_ZqpQp}\\
q+\bar q&\to&Z+g+g,\label{qQ_Zgg}\\
q+\bar q&\to&Z+g+\gamma,\label{qQ_Zgy}\\
q+q&\to&Z+q+q,\label{qq_Zqq}\\
q+q^\prime&\to&Z+q+q^\prime,\label{qqp_Zqqp}\\
q+g&\to&Z+q+g,\label{qg_Zqg}\\
q+g&\to&Z+q+\gamma,\label{qg_Zqy}\\
g+g&\to&Z+q+\bar q,\label{gg_ZqQ}
\end{eqnarray}
where $q^\prime\ne q$.
Subprocesses (\ref{qQ_ZqQ}) and (\ref{qq_Zqq}) contribute both at
$\mathcal{O}(\alpha\alpha_s^2)$ and $\mathcal{O}(\alpha^2\alpha_s)$,
subprocesses (\ref{qQ_ZqpQp}), (\ref{qQ_Zgg}), (\ref{qqp_Zqqp}),
(\ref{qg_Zqg}), and (\ref{gg_ZqQ}) contribute at
$\mathcal{O}(\alpha\alpha_s^2)$, and subprocesses (\ref{qQ_Zgy}) and
(\ref{qg_Zqy}) contribute at $\mathcal{O}(\alpha^2\alpha_s)$.
The tree-level diagrams contributing to subprocess (\ref{qQ_ZqQ}) are shown in
Fig.~\ref{fig:qQZqQ}, those contributing to subprocesses (\ref{qQ_Zgg}) and
(\ref{qQ_Zgy}) in Fig.~\ref{fig:qQZgg}, and those contributing to subprocesses
(\ref{qq_Zqq}) and (\ref{qqp_Zqqp}) in Fig.~\ref{fig:qqZqq}.

%
%  Fig. 1
%
%\vspace*{2.0cm}
%\SetScale{0.8}
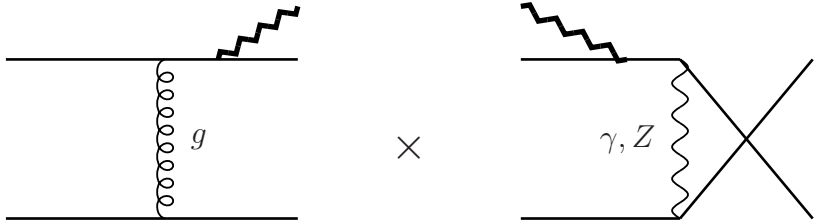
\begin {figure} [htbp]
~~~~~~~
\begin{picture}(150,100)(0,0)
\SetWidth{1.0}
\Line(20,20)(130,20)
\Line(20,80)(130,80)
\SetWidth{0.7}
\Gluon(80,20)(80,80){3}{8}
%\Photon(80,20)(80,80){3}{5}
\SetWidth{2.0}
\ZigZag(100,80)(130,100){2}{4}
\Text(90,50)[l]{$g$}
\end{picture}
~~
\raisebox{15mm}{\Large $\times$}
~~
\begin{picture}(150,100)(0,0)
\SetWidth{1.0}
\Line(20,20)(80,20)
\Line(20,80)(80,80)
\Line(80,80)(130,20)
\Line(80,20)(130,80)
\SetWidth{0.7}
%\Gluon(80,20)(80,80){3}{8}
\Photon(80,20)(80,80){3}{5}
\SetWidth{2.0}
\ZigZag(60,80)(20,100){2}{4}
\Text(50,50)[l]{$\gamma,Z$}
\end{picture}
\caption{\label{fig:color}%
$\mathcal{O}(\alpha^{1/2}\alpha_s)$ and $\mathcal{O}(\alpha^{3/2})$
tree-level diagrams interfering to yield $\mathcal{O}(\alpha^2\alpha_s)$
contributions to subprocesses $q+\bar q\to Z+q+\bar q$ and $q+q\to Z+q+q$.}
\label{figcolor}
\end{figure}

As already mentioned in Sec.~\ref{sec:one}, the $\mathcal{O}(\alpha^2\alpha_s)$
contributions to subprocesses (\ref{qQ_ZqQ}) and (\ref{qq_Zqq}) are generated
by interferences of $2\to3$ tree-level diagrams with a virtual gluon in one
factor and a virtual photon or $Z$ boson in the other one in such a way that
one closed quark line is formed yielding a nonvanishing color factor,
$\tr T^aT^a=N_cC_F$, as indicated in Fig.~\ref{fig:color}.
In the case of subprocess (\ref{qQ_ZqQ}), this is achieved when the gluon is
in the $s$ channel and the photon or $Z$ boson is in the $t$ channel or vice
versa.
In the case of subprocess (\ref{qq_Zqq}), this is achieved by twisting the
quark lines in one of the interfering diagrams.
These interference contributions exhaust the $\mathcal{O}(\alpha^2\alpha_s)$
corrections to subprocesses (\ref{qQ_ZqQ}) and (\ref{qq_Zqq}) and are thus
finite and gauge-independent by themselves.
On the other hand, interferences leading to two closed quark lines are
nullified by $(\tr T^a)^2=0$.
This explains, why subprocesses (\ref{qQ_ZqpQp}) and (\ref{qqp_Zqqp}) do not
receive $\mathcal{O}(\alpha^2\alpha_s)$ contributions.
Obviously, these types of mixed QCD-QED corrections may not be obtained by
merely manipulating coupling constants and color factors as is often the case
for pure QED corrections.

We compute the full $\mathcal{O}(\alpha\alpha_s^2)$ and
$\mathcal{O}(\alpha^2\alpha_s)$ corrections to the cross section of the
hadronic process (\ref{eq:hadron}) according to Eq.~(\ref{eq:xs}) by including
all the partonic subprocesses (\ref{qQ_Zg})--(\ref{gg_ZqQ}).
We regularize both the UV and IR divergences using dimensional regularization
in $D=4-2\varepsilon$ space-time dimensions.
The UV divergences arise from the $2\to2$ one-loop diagrams and are removed by
renormalizing the coupling constants, masses, and wave functions in the
respective $2\to2$ tree-level diagrams. 
We renormalize $\alpha_s$ and $\alpha$ according to the modified
minimal-subtraction ($\overline{\mathrm{MS}}$) scheme and employ the
electroweak on-shell renormalization scheme otherwise.
In particular, we define $\theta_w$ in terms of the pole masses as
$\cos\theta_w=M_W/M_Z$.
The IR divergences, both of soft and collinear types, are generated by $2\to2$
one-loop and $2\to3$ tree-level diagrams.
The soft and collinear divergences related to the final states are canceled by
integrating over the kinematic degrees of freedom of the three-particle phase
space that are related to the systems $X$ and combining the outcome with the
virtual corrections.
Specifically, the three-particle kinematics turns into the two-particle one by
taking the limit $s_2\to0$.
This is implemented in dimensional regularization using the relationship
\begin{equation}
\frac{1}{s_2^{1+\varepsilon}} = 
  \delta(s_2)\left( - \frac{1}{\varepsilon} + \ln{s_2^{\rm max}}
    -\frac{\varepsilon}{2}\ln^2{s_2^{\rm max}} + \cdots \right)   
  + \left( \frac{1}{s_2} \right)_{+}
  + \left( \frac{\ln s_2}{s_2} \right)_{+}  
  + \cdots ,
\label{eq:s2}
\end{equation}
where $s_2^{\rm max}$ is the maximum value of $s_2$ allowed for given values of
$p_T$ and $y$ and the plus distributions are defined for smooth test functions
$f(s_2)$ as
\begin{eqnarray}
\int_0^{s_2^{\rm max}} ds_2\left(\frac{1}{s_2}\right)_+ f(s_2) &=& 
   \int_0^{s_2^{\rm max}}ds_2\frac{1}{s_2}[f(s_2)-f(0)], \nonumber\\
\int_0^{s_2^{\rm max}} ds_2\left(\frac{\ln s_2}{s_2}\right)_+ f(s_2) &=& 
   \int_0^{s_2^{\rm max}}ds_2\frac{\ln s_2}{s_2}[f(s_2)-f(0)].
\end{eqnarray}

There remain collinear divergences related to the initial states, which are
universal and are absorbed into the bare PDFs so to render them finite.
In the $\overline{\mathrm{MS}}$ factorization scheme, this PDF renormalization
is implemented in the QCD sector as
\begin{equation}
  f_{i/h}(x,\mu_F^2) = \sum_j\int\limits_x^1 \frac{dy}{y}
    f_{j/h}^{\rm bare}\left(\frac{x}{y}\right) \left[ \delta_{ij}\delta(1-y) 
     - \frac{\mu_F^{-2\varepsilon}}{\varepsilon}\, \frac{\alpha_s}{2\pi}\, 
      \frac{\Gamma(1-\varepsilon)}{\Gamma(1-2\varepsilon)} P_{ij}(y)
     + \cdots \right],
\label{eq:fact}
\end{equation}
where $P_{ij}(y)$ are the $j\to i$ splitting functions.
In the one-loop approximation of QCD, the latter read \cite{Altarelli:1977zs}
\begin{eqnarray}
  P_{qq}(y) &=& C_F \left[\frac{3}{2}\delta(1-y)
+2\left(\frac{1}{1-y}\right)_+-1-y\right] ,
\nonumber\\
  P_{gq}(y) &=& C_F \frac{1+(1-y)^2}{y},
\nonumber \\
  P_{gg}(y) &=&\left(\frac{11}{6}C_A-\frac{2}{3}Tn_f\right)\delta(1-y)
+2C_A\left[\left(\frac{1}{1-y}\right)_++\frac{1}{y}-2 + y(1-y)\right],
\nonumber\\
  P_{qg}(y) &=& T[y^2+(1-y)^2],
\label{eq:splitting}
\end{eqnarray}
where $C_F=(N_c^2-1)/(2N_c)=4/3$, $C_A=N_c=3$, $T=1/2$, and $n_f=5$ is the
number of active quark flavors.
For simplicity, we adopt the $\overline{\mathrm{MS}}$ factorization scheme also
for the QED sector.
The appropriate counterparts of Eqs.~(\ref{eq:fact}) and (\ref{eq:splitting})
are obtained by substituting $\alpha_s\to\alpha Q_q^2$, $C_F\to1$, and
$C_A\to0$.
Of course, the PDFs to be used in the numerical analysis must then be
implemented with the $\overline{\mathrm{MS}}$ factorization scheme as well,
both in their QCD and QED sectors.
While the $\overline{\mathrm{MS}}$ factorization scheme is now common standard
for the QCD sector, alternative choices have been advocated for the QED sector,
e.g.\ in connection with neutrino-nucleus deep-inelastic scattering (DIS), a
DIS-like choice \cite{Diener:2005me}.

To exploit Eq.~(\ref{eq:s2}), it is useful to introduce $s_2$ as an integration
variable in Eq.~(\ref{eq:xs}), in lieu of $x_2$, say.
This leads to
\begin{eqnarray}
  \frac{d\sigma}{dq_T^2\,dy} = \sum_{i,j} \int\limits_{x_1^{\rm min}}^1dx_1
     \int\limits_0^{s_2^{\rm max}} \!\! \frac{ds_2}{x_1 S+U-Q^2} 
     f_{i/h_1}(x_1,\mu_F^2) f_{j/h_2}(x_2,\mu_F^2)
     \frac{s\,d\sigma_{ij}}{dt\,du}(x_1 P_1, x_2 P_2, \mu_F^2),\nonumber\\
&&
\end{eqnarray}
where
\begin{eqnarray}
x_1^{\rm min} &=& \frac{-U}{S+T-Q^2},\qquad
s_2^{\rm max} = U + x_1(S+T-Q^2),\qquad
x_2 = \frac{s_2-Q^2-x_1(T-Q^2)}{x_1 S+U-Q^2},\nonumber\\
T &=& Q^2 - e^{-y}\sqrt{S(Q^2+q_T^2)}, \qquad
U = Q^2 - e^y\sqrt{S(Q^2+q_T^2)}.
\label{eq:alt}
\end{eqnarray}

At this point, we compare our analytic results with the literature
\cite{Gonsalves:1989ar,Kuhn:2005az}.
The NLO QCD corrections, of relative order $\mathcal{O}(\alpha_s)$, due to the
virtual contributions from subprocesses (\ref{qQ_Zg}) and (\ref{qg_Zq}), and
the real contributions from subprocesses
%(\ref{qQ_ZqQ}), (\ref{qQ_ZqpQp}), (\ref{qQ_Zgg}),
(\ref{qQ_ZqQ})--(\ref{qQ_Zgg}),
%(\ref{qq_Zqq}), (\ref{qqp_Zqqp}), (\ref{qg_Zqg}),
(\ref{qq_Zqq})--(\ref{qg_Zqg}),
and (\ref{gg_ZqQ}) are listed in Ref.~\cite{Gonsalves:1989ar}.
Apart from some misprints,\footnote{%
In Eq.~(2.12) of Ref.~\cite{Gonsalves:1989ar}, $B_2^{qG}(s,t,u,Q^2)$ should be
replaced with $\left[B_2^{qG}(s,t,u,Q^2)+C_2^{qG}(s,t,u,Q^2)\right]$ and
$C_2^{qG}(s,t,u,Q^2)$ with $C_3^{qG}(s,t,u,Q^2)$.}
we find agreement with Ref.~\cite{Gonsalves:1989ar}.
The weak $\mathcal{O}(\alpha)$ corrections to subprocess (\ref{qQ_Zg}) are
listed in Ref.~\cite{Kuhn:2005az}.
In Ref.~\cite{Kuhn:2005az}, collinear divergences arising from box diagrams in
intermediate steps of the calculation are regularized by introducing an
infinitesimal quark mass $\lambda$, while we employ dimensional regularization.
The $\lambda$-dependent one-loop scalar box integrals $J_{12}$, $J_{13}$, and
$J_{14}$ in Eq.~(39) of Ref.~\cite{Kuhn:2005az} may be conveniently converted to
dimensional regularization using the results of Ref.~\cite{Ellis:2007qk}.
In Ref.~\cite{Kuhn:2005az}, the renormalization is performed both in the
$\overline{\mathrm{MS}}$ scheme and in the on-shell scheme implemented with
some running fine-structure constant as explained in Eqs.~(49) and (50) of
Ref.~\cite{Kuhn:2005az}, which differs from the pure $\overline{\mathrm{MS}}$
definition.
Specifically, in the counterterm of the electromagnetic coupling constant, the
photon self-energy, which appears there with argument $q^2=0$ in the pure
on-shell scheme, is split into the fermionic and bosonic parts, and the
argument of the former is shifted to $q^2=M_Z^2$.
While the latter construction is well defined at one loop, it becomes ambiguous
at higher orders because of the required separation of fermionic and bosonic
contributions, and it is bound to render the running fine-structure constant
thus defined gauge dependent.
By contrast, we work in a hybrid renormalization scheme, which uses the pure
$\overline{\mathrm{MS}}$ definition of $\alpha$, but the electroweak on-shell
scheme otherwise.
Taking these conceptual differences into account, we fully agree with
Ref.~\cite{Kuhn:2005az}.

In this paper, we only list those analytic results that may not be found in the
previous literature.
Specifically, we consider $\mathcal{O}(\alpha^2\alpha_s)$ contributions to
$q+\bar q\to Z+X$, $q+g\to Z+X$, and $q+q\to Z+X$ in Appendices~\ref{sec:a},
\ref{sec:b}, and \ref{sec:c}, respectively.
In the case of $q+\bar q\to Z+X$, this includes the virtual QED corrections to
subprocess (\ref{qQ_Zg}), the virtual QCD corrections to subprocess
(\ref{qQ_Zy}), the real corrections from subprocess (\ref{qQ_Zgy}), and the
above-mentioned interference contributions from subprocess (\ref{qQ_ZqQ}).
In the case of $q+g\to Z+X$, this includes the virtual QED corrections to
subprocess (\ref{qg_Zq}) and the real corrections from subprocess
(\ref{qg_Zqy}).
In the case of $q+q\to Z+X$, this includes the above-mentioned interference
contributions from subprocess (\ref{qq_Zqq}).

As already mentioned in Sec.~\ref{sec:one}, we also include the LO
contributions from photoproduction.
Incoming photons can participate in the hard scattering either directly or
indirectly, i.e.\ through their quark and gluon content.
The contributions from direct and resolved photoproduction are formally of the
same order in the perturbative expansion.
This may be understood by observing that the PDFs of the photon have a leading
behavior proportional to
$\alpha\ln(\mu_F^2/\Lambda_\mathrm{QCD}^2)\propto\alpha/\alpha_s(\mu_F^2)$,
where $\Lambda_\mathrm{QCD}$ is the asymptotic scale parameter of QCD.
At LO, direct photoproduction proceeds via subprocess (\ref{qy_Zq}) and
resolved photoproduction via subprocesses (\ref{qQ_Zg}) and (\ref{qg_Zq}).
The cross section of subprocess (\ref{qy_Zq}) reads \cite{Brown:1975yc}
\begin{equation}
\frac{d\sigma_{q\gamma}}{dt}=-\frac{2\pi\alpha^2Q_q^2(v_q^2+a_q^2)}{N_cs^2}
\left(\frac{t}{s}+\frac{s}{t}+\frac{2uQ^2}{st}\right),
\end{equation}
where the Mandelstam variables and gauge coupling constants are defined in
Eqs.~(\ref{eq:mandelstam}) and (\ref{eq:Zqq}), respectively.
The ones of subprocesses (\ref{qQ_Zg}) and (\ref{qg_Zq}) may be read off from
Eqs.~(\ref{eq:qQ}) and (\ref{eq:qg}), respectively.
The emission of photons off the (anti)proton can happen either elastically or
inelastically, i.e.\ the (anti)proton stays intact or is destroyed,
respectively.
In both cases, an appropriate PDF can be evaluated in the
Weizs\"acker-Williams approximation
\cite{Kniehl:1990iv,Gluck:1994vy,Martin:2004dh,Ball:2013hta}.
Since these PDFs are of $\mathcal{O}(\alpha)$, the LO photoproduction
contributions are of $\mathcal{O}(\alpha^3)$.
Although photoproduction contributions are parametrically suppressed by a
factor of $\alpha/\alpha_s$ relative to the $\mathcal{O}(\alpha^2\alpha_s)$
corrections discussed above, we include them in our analysis because they may
turn out to be sizable in certain regions of phase space.

We generated the Feynman diagrams using the program package DIANA
\cite{Tentyukov:1999is} and checked the output using the program package
FeynArts~3 \cite{Hahn:2000kx}.
We reduced the one-loop tensor integrals to scalar ones using custom-made
routines written with the symbolic manipulation program FORM version~4.0
\cite{Kuipers:2012rf}.
We evaluated the scalar one-loop integrals using the analytic results listed in
Ref.~\cite{Ellis:2007qk}.

\section{Numerical analysis}
\label{sec:three}

We are now in a position to present our numerical analysis.
As input we use the pole masses
$M_W=80.385$~GeV,
$M_Z=91.1876$~GeV,
$M_H=125$~GeV,
$m_b=4.89$~GeV, and
$m_t=173.07$~GeV, and
the $\overline{\mathrm{MS}}$ coupling constants $\bar\alpha(M_Z^2)=1/127.944$
\cite{Beringer:1900zz} and $\alpha_s^{(5)}(M_Z^2)=0.1180$ \cite{Ball:2013hta}
to gauge $\bar\alpha(\mu_R^2)$ and $\alpha_s^{(5)}(\mu_R^2)$.
We employ the NNPDF2.3QED NLO set of proton PDFs \cite{Ball:2013hta}, which
also include QED evolution and provide a photon distribution function.
This allows us to consistently treat direct and resolved photoproduction via
inelastic scattering off the (anti)proton along with ordinary hadroproduction.
In Ref.~\cite{Ball:2013hta}, the QED sector is treated at LO, or, more
accurately, at the leading logarithmic level, where factorization is still
trivial and does not yet require the specification of a scheme.
In this sense, the NNPDF2.3QED NLO PDFs are compatible with our convention of
employing the $\overline{\mathrm{MS}}$ factorization scheme in the QED sector
\cite{Frixione:2015zaa}.
By the same token, the dependence on the QED factorization scheme contributes
to the theoretical uncertainty, which we refrain from assessing here. 
As for photoproduction via elastic scattering off the (anti)proton, we adopt
the photon flux function from Ref.~\cite{Kniehl:1990iv} and the resolved-photon
PDFs from Ref.~\cite{Aurenche:2005da}.
For definiteness, we identify the renormalization and factorization scales with
the $Z$-boson transverse mass, $\mu_R=\mu_F=\sqrt{p_T^2+M_Z^2}$.

\begin{figure}
\begin{center}
\includegraphics*[width=0.7\linewidth,angle=0]{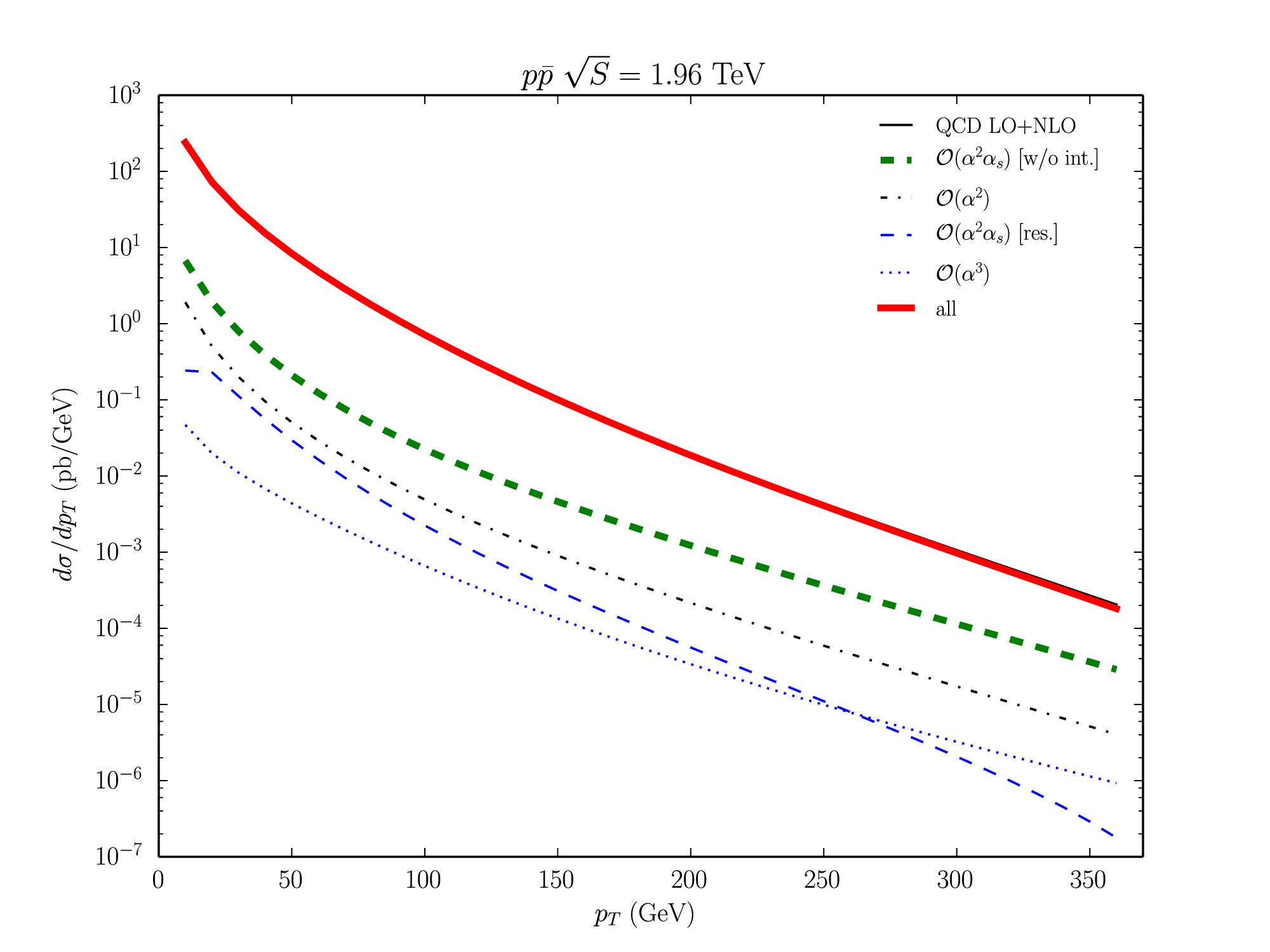}
\leftline{(a)}
\includegraphics*[width=0.7\linewidth,angle=0]{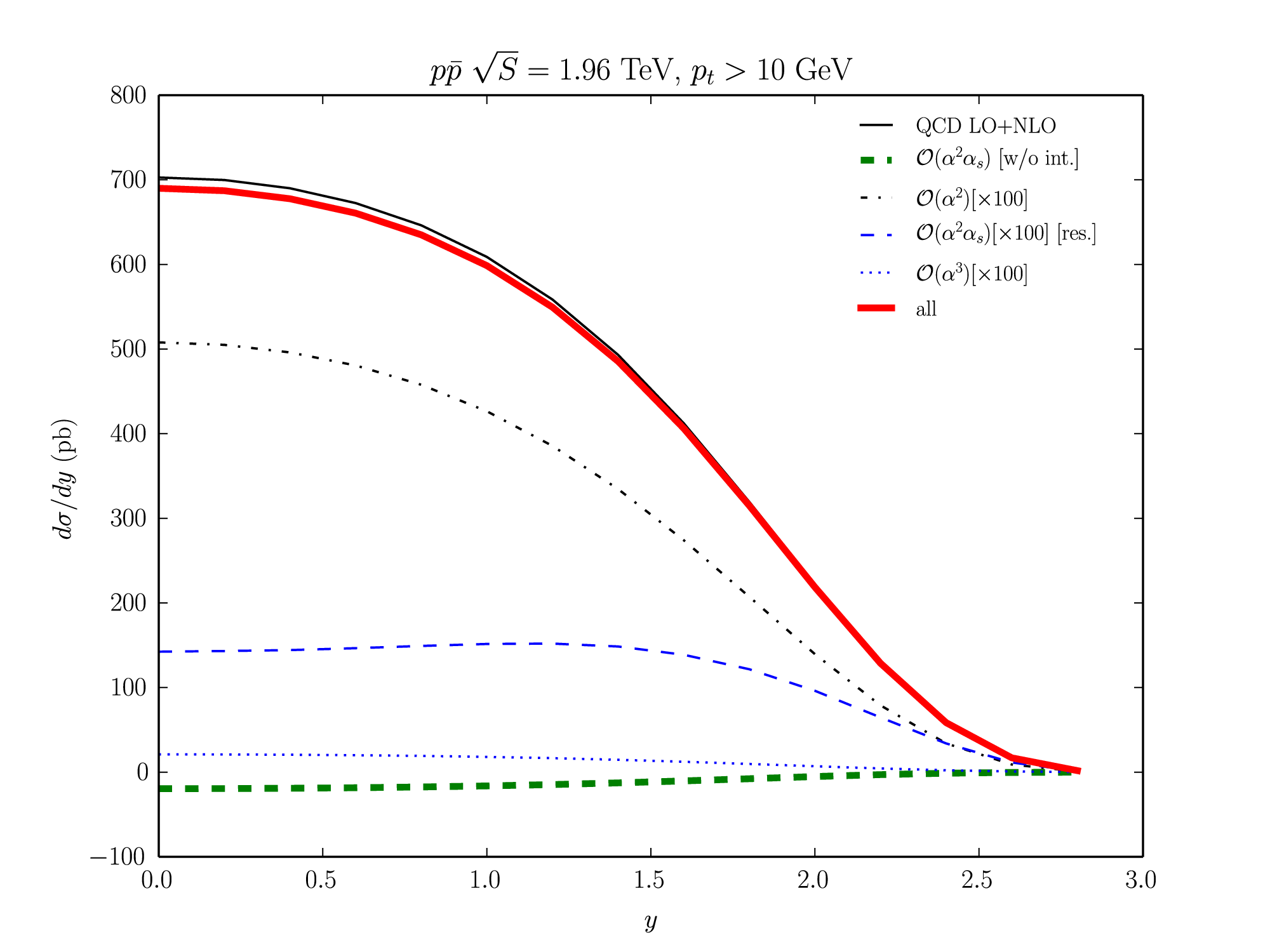}
\leftline{(b)}
\end{center}
\caption{\label{fig:tevatron}%
Cross section distributions in (a) $p_T$ and (b) $y$ for $p_T>10$~GeV of
$p\bar p\to Z+X$ at $\sqrt S=1.96$~TeV (Tevatron run~II).
In each frame, the NLO QCD result \cite{Gonsalves:1989ar}, i.e.\ the sum of
the $\mathcal{O}(\alpha\alpha_s)$ and $\mathcal{O}(\alpha\alpha_s^2)$ results
(thin solid lines),
the $\mathcal{O}(\alpha^2)$ Born result (thin dot-dashed lines),
the purely weak $\mathcal{O}(\alpha^2\alpha_s)$ corrections to
subprocesses (\ref{qQ_Zg}) and (\ref{qg_Zq}) \cite{Kuhn:2005az} (thick dashed
green lines),
the residual electroweak $\mathcal{O}(\alpha^2\alpha_s)$ corrections (thin
dashed blue lines),
the $\mathcal{O}(\alpha^3)$ photoproduction contributions (thin dotted blue
lines),
and the total sum (thick solid red lines) are shown.} 
\end{figure}

\begin{figure}
\begin{center}
\includegraphics*[width=0.7\linewidth,angle=0]{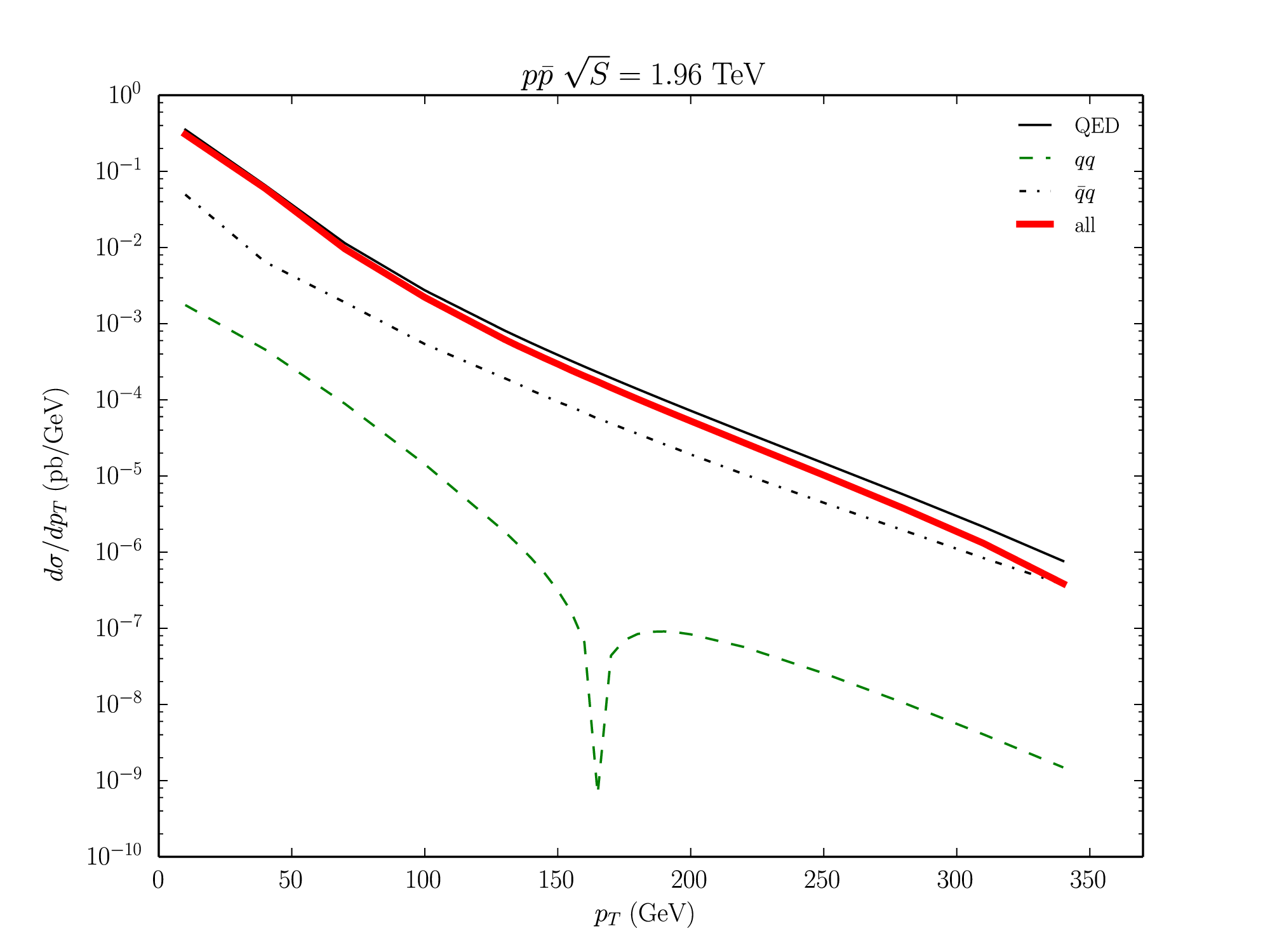}
\leftline{(a)}
\includegraphics*[width=0.7\linewidth,angle=0]{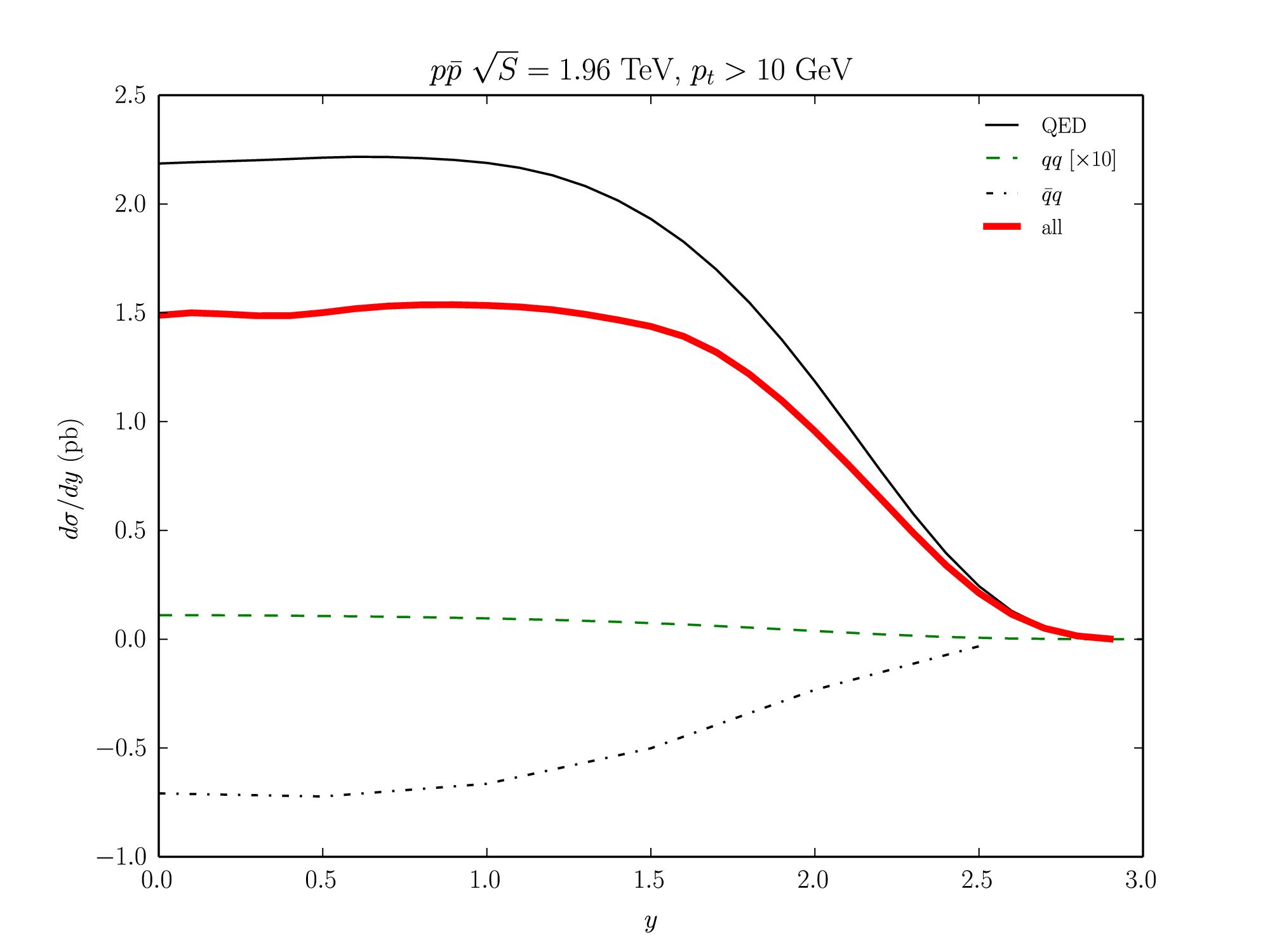}
\leftline{(b)}
\end{center}
\caption{\label{fig:tevatron1}%
The residual electroweak $\mathcal{O}(\alpha^2\alpha_s)$ corrections in
Fig.~\ref{fig:tevatron} (thick solid red lines) are decomposed into the
combination of the $\mathcal{O}(\alpha)$ QED corrections to subprocesses
(\ref{qQ_Zg}) and (\ref{qg_Zq}) and the $\mathcal{O}(\alpha_s)$ QCD corrections
to subprocess (\ref{qQ_Zy}) (thin solid lines), and
the QCD-electroweak interference contributions from subprocesses
(\ref{qQ_ZqQ}) (thin dot-dashed lines) and (\ref{qq_Zqq}) (thin dashed green
lines).}
\end{figure}

\begin{figure}
\begin{center}
\includegraphics*[width=0.7\linewidth,angle=0]{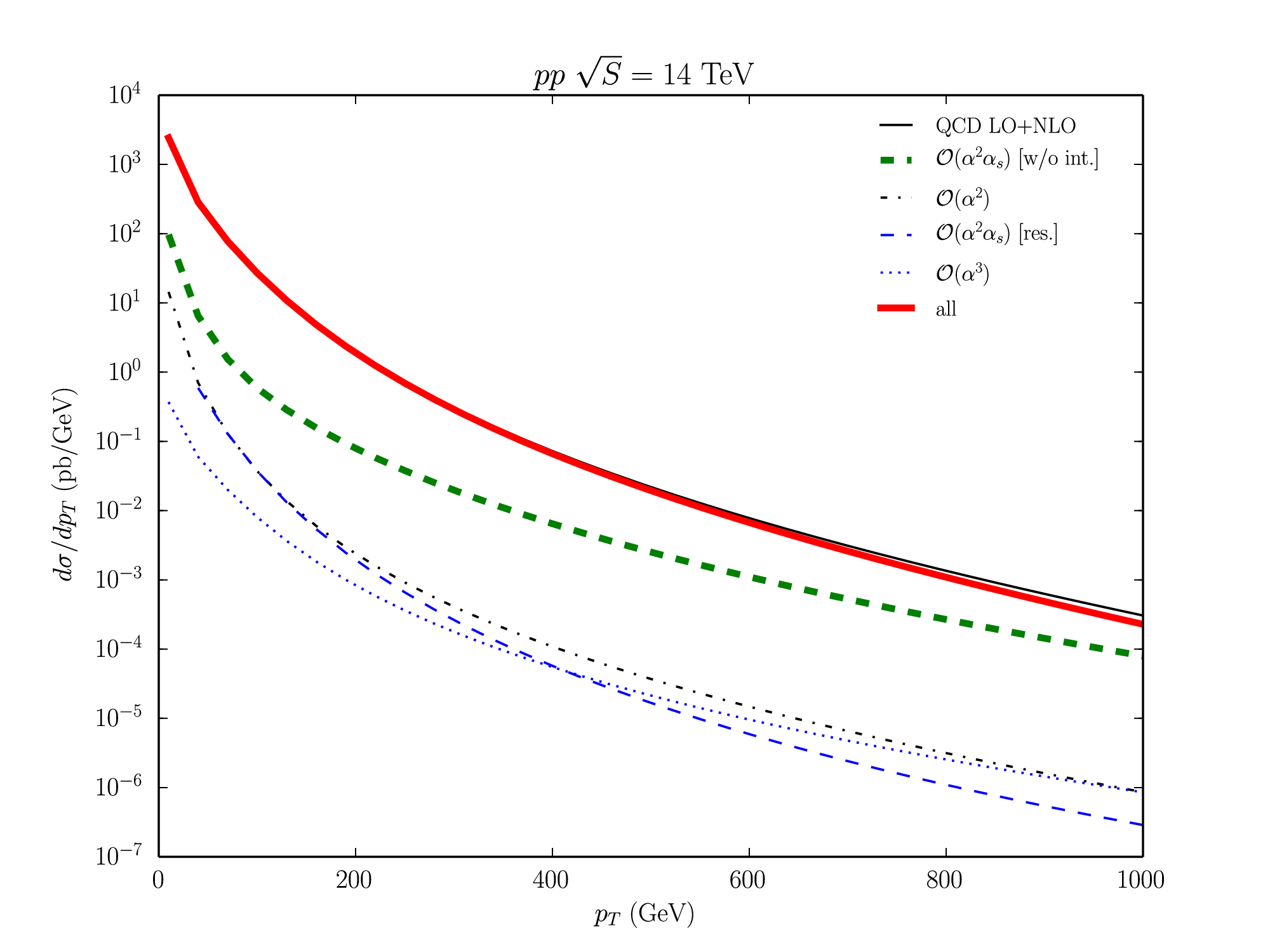}
\leftline{(a)}
\includegraphics*[width=0.7\linewidth,angle=0]{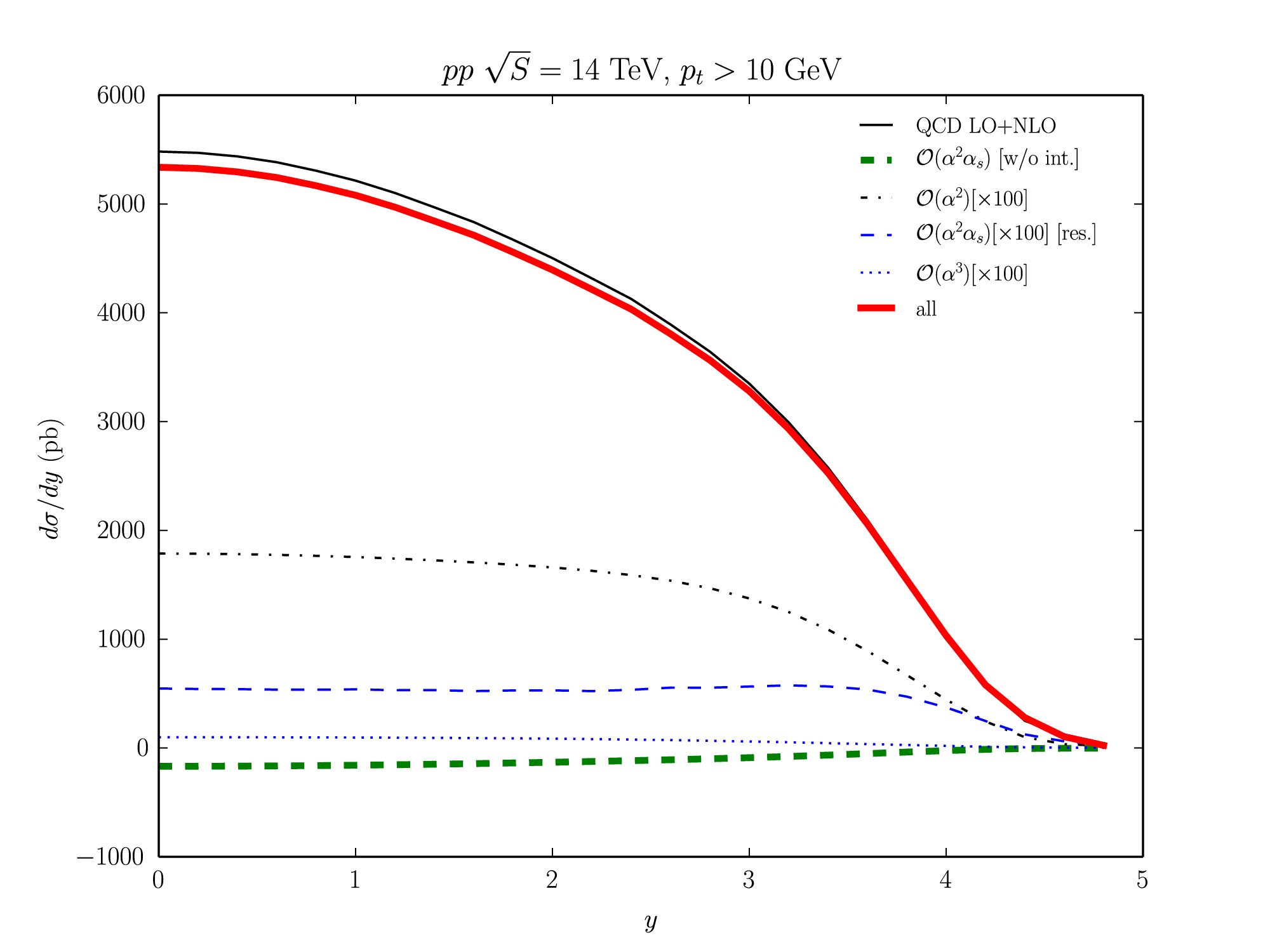}
\leftline{(b)}
\end{center}
\caption{\label{fig:lhc}%
Same as in Fig.~\ref{fig:tevatron}, but for $pp\to Z+X$ at $\sqrt S=14$~TeV
(LHC).} 
\end{figure}

\begin{figure}
\begin{center}
\includegraphics*[width=0.7\linewidth,angle=0]{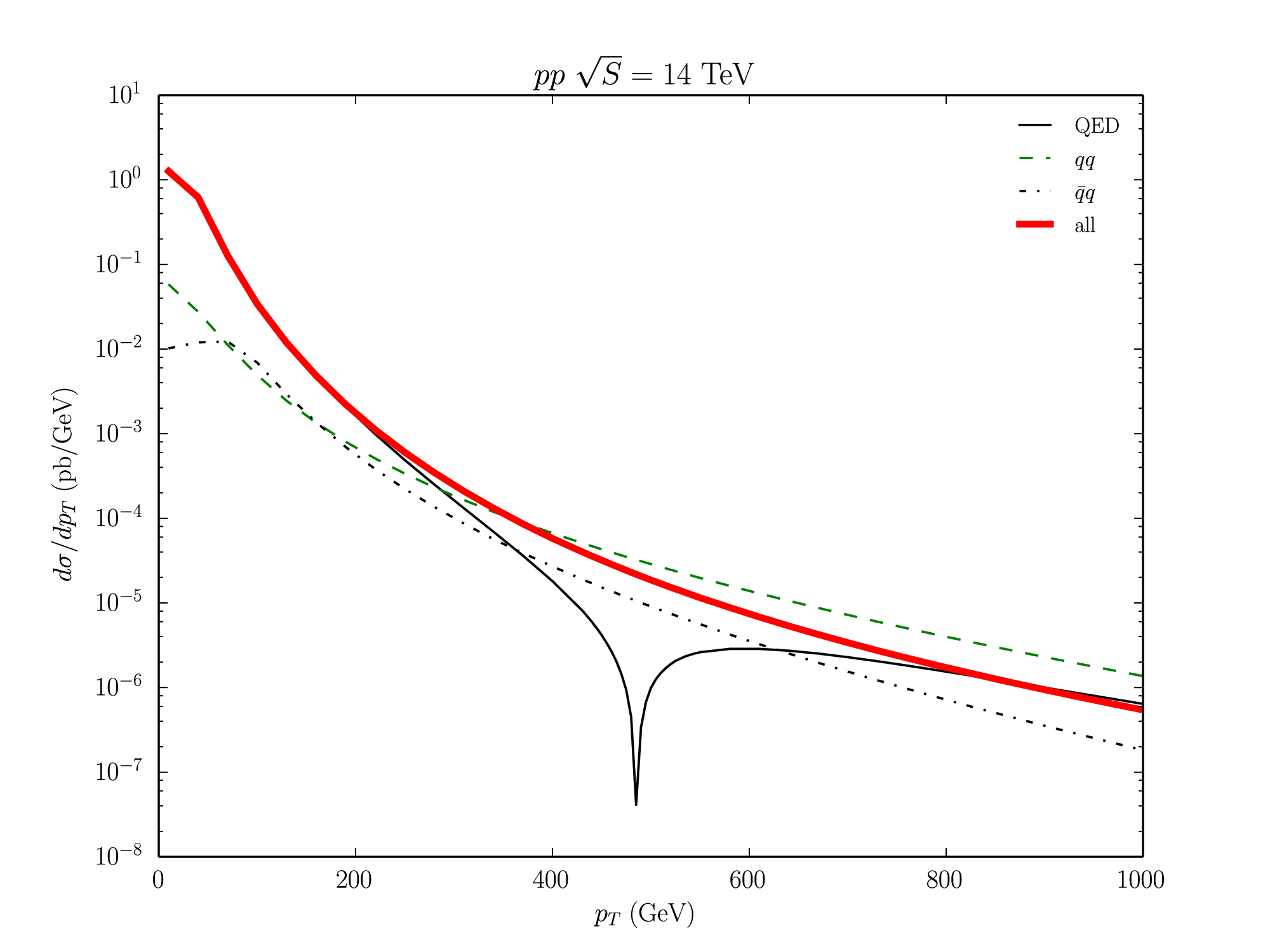}
\leftline{(a)}
\includegraphics*[width=0.7\linewidth,angle=0]{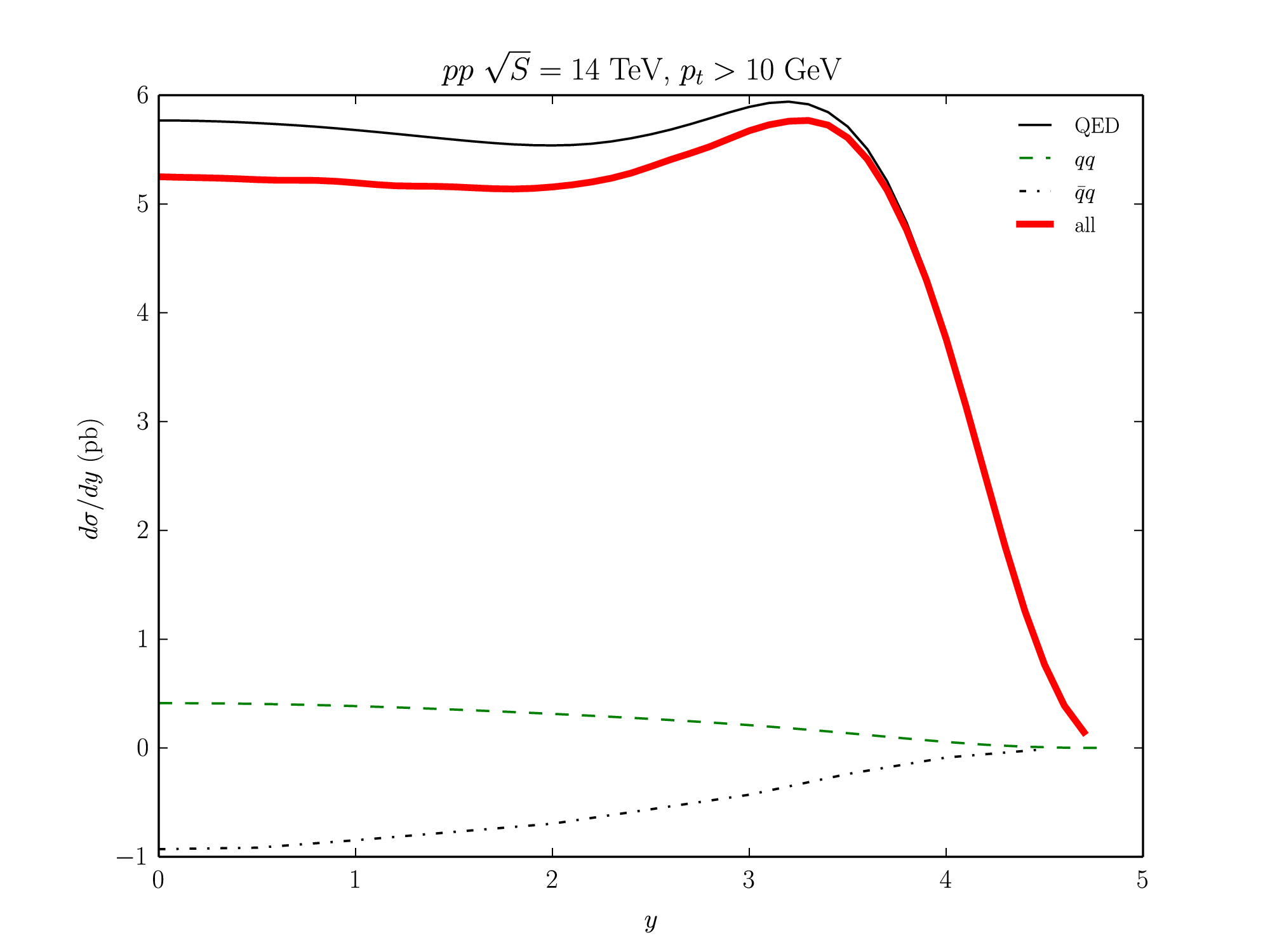}
\leftline{(b)}
\end{center}
\caption{\label{fig:lhc1}%
Same as in Fig.~\ref{fig:tevatron1}, but for $pp\to Z+X$ at $\sqrt S=14$~TeV
(LHC).} 
\end{figure}

In Fig.~\ref{fig:tevatron}, we study the cross section of $p\bar p\to Z+X$ at
center-of-mass energy $\sqrt S=1.96$~TeV appropriate for Tevatron run~II (a)
differential in $p_T$ integrated over $y$ and (b) differential in $y$ imposing
the acceptance cut $p_T>10$~GeV.
Specifically, we show
(i) the NLO QCD result considered in Ref.~\cite{Gonsalves:1989ar}, i.e.\ the
sum of the $\mathcal{O}(\alpha\alpha_s)$ and $\mathcal{O}(\alpha\alpha_s^2)$
results (thin solid lines);
(ii) the $\mathcal{O}(\alpha^2)$ Born result (thin dot-dashed lines);
(iii) the purely weak $\mathcal{O}(\alpha^2\alpha_s)$ corrections to
subprocesses (\ref{qQ_Zg}) and (\ref{qg_Zq}) considered in
Ref.~\cite{Kuhn:2005az} (thick dashed green lines);
(iv) the residual electroweak $\mathcal{O}(\alpha^2\alpha_s)$ corrections
(thin dashed blue lines);
(v) the $\mathcal{O}(\alpha^3)$ photoproduction contributions (thin dotted blue
lines);
and (vi) the total sum (thick solid red lines).
The $p_T$ distributions in Fig.~\ref{fig:tevatron}(a) are plotted on a
logarithmic scale.
Since the one of contribution~(iii) is negative in the considered $p_T$ range,
its modulus is shown.
The $y$ distributions in Fig.~\ref{fig:tevatron}(b) are plotted on a linear
scale.
For better visibility, contributions~(ii), (iv), and (v) are amplified by a
factor of 100.
In Fig.~\ref{fig:tevatron}(b), we do not consider negative $y$ values because
the $y$ distributions are symmetric by charge conjugation invariance.
In Fig.~\ref{fig:tevatron1}, we decompose contribution~(iv) (thick solid red
lines) into the combination of the $\mathcal{O}(\alpha)$ QED corrections to
subprocesses (\ref{qQ_Zg}) and (\ref{qg_Zq}) and the $\mathcal{O}(\alpha_s)$
QCD corrections to subprocess (\ref{qQ_Zy}) (thin solid lines), which cannot be
usefully separated, and the QCD-electroweak interference contributions from
subprocesses (\ref{qQ_ZqQ}) (thin dot-dashed lines) and (\ref{qq_Zqq}) (thin
dashed green lines). 
In Figs.~\ref{fig:lhc} and \ref{fig:lhc1}, we repeat the analyses of
Figs.~\ref{fig:tevatron} and \ref{fig:tevatron1}, respectively, for
$pp\to Z+X$ at $\sqrt S=14$~TeV appropriate for the LHC.

From Figs.~\ref{fig:tevatron} and \ref{fig:lhc}, we observe that the combined
effect of the electroweak contributions (ii)--(v) is to reduce the NLO QCD
predictions (i).
The reduction ranges from a few percent at low $p_T$ values to a few tens of
percent in the large-$p_T$ domain, where large Sudakov logarithms dominate.
The bulk of the electroweak contributions (ii)--(v) is made up by the purely
weak $\mathcal{O}(\alpha^2\alpha_s)$ corrections to subprocesses (\ref{qQ_Zg})
and (\ref{qg_Zq}) [contribution (iii)], which are negative throughout the
kinematic ranges considered here.
The residual types of electroweak effects taken into accout here, namely the
$\mathcal{O}(\alpha^2)$ Born result (ii),
the residual electroweak $\mathcal{O}(\alpha^2\alpha_s)$ corrections (iv), and
the $\mathcal{O}(\alpha^3)$ photoproduction contributions (v), are all
positive, but numerically suppressed by typically one order of magnitude or
more relative to contribution~(iii), except for contribution~(ii) in the
small-$p_T$ range.

In Fig.~\ref{fig:tevatron1}(a), the QED-type correction is throughout
positive, the $q\bar{q}$ interference contribution is throughout negative, and
the $qq$ interference contribution is negative for $p_T\agt165$~GeV.
On the other hand, in Fig.~\ref{fig:lhc1}(a), the QED-type correction is
negative for $p_T\agt480$~GeV, the $q\bar{q}$ interference contribution is
again throughout negative, and the $qq$ interference contribution is
throughout positive.
From Figs.~\ref{fig:tevatron1}(a) and \ref{fig:lhc1}(a), we observe that those
two interference contributions are suppressed relative to the QED-type
correction in the small-$p_T$ range.
This is also reflected in the $y$ distributions of Figs.~\ref{fig:tevatron1}(b)
and \ref{fig:lhc1}(b), which receive dominant contributions from the
small-$p_T$ ranges.
Such a suppression is expected from the comparison of color factors
\cite{Denner:2011vu,Denner:2012ts}.
However, the situation is quite different at large $p_T$ values.
In fact, in Fig.~\ref{fig:tevatron1}(a), the $q\bar{q}$ interference
contribution steadily approaches the QED-type contribution for increasing value
of $p_T$, and, in Fig.~\ref{fig:lhc1}(a), the $qq$ interference contribution
exceeds the QED-type contribution for $p_T\agt300$~GeV.
Comparing Figs.~\ref{fig:tevatron1} and \ref{fig:lhc1} with
Figs.~\ref{fig:tevatron} and \ref{fig:lhc}, we observe that the three
$\mathcal{O}(\alpha^2\alpha_s)$ contributions of class~(iv) range at the
permille level with respect to the well-known NLO QCD result
\cite{Gonsalves:1989ar}.
Specifically, in the case of $\sigma/dy$ at $y=0$, the QED-type, $qq$
interference, and $q\bar{q}$ interference contributions normalized to the NLO
QCD result approximately amount
to $3\times10^{-3}$, $2\times10^{-4}$, and $-1\times10^{-4}$ at the Tevatron and
to $1\times10^{-3}$, $8\times10^{-5}$, and $-2\times10^{-4}$ at the LHC,
respectively.

\section{Conclusions}
\label{sec:four}

We studied the inclusive hadroproduction of large-$p_T$ single $Z$ bosons
including both the QCD and electroweak NLO corrections and presented $p_T$ and
$y$ distributions under Tevatron and LHC kinematic conditions.
Our analytic results agree with the literature as far as the latter goes.
Specifically, we recovered the well-known NLO QCD corrections
\cite{Gonsalves:1989ar}, of absolute order $\mathcal{O}(\alpha\alpha_s^2)$, and
the purely weak one-loop corrections, of absolute order
$\mathcal{O}(\alpha^2\alpha_s)$, to the partonic subprocess $q\bar q\to Zg$ and
its crossed versions \cite{Kuhn:2005az}.
We completed our knowledge of the $\mathcal{O}(\alpha^2\alpha_s)$ corrections
by providing also the QED corrections and some missing weak contributions due
to interferences of tree-level $2\to3$ scattering amplitudes in compact
analytic form ready to be used by the interested reader.
While the new $\mathcal{O}(\alpha^2\alpha_s)$ contributions turned out to be
numerically small, their knowledge should help us to reduce the theoretical
uncertainty on this important benchmark cross section.
We also considered, for the first time, direct and resolved photoproduction in
elastic and inelastic scattering.

In the experimental analyses to be compared with our theoretical predictions,
$Z$ bosons must be identified, preferrably via their decays to $e^+e^-$ or
$\mu^+\mu^-$ pairs, and their four-momenta must be reconstructed and sampled in
bins of $p_T$ or $y$, ignoring any other available information about the
selected events.
If there are more than one identified and reconstructed $Z$ boson in an event,
then each of them is taken to generate one entry in the considered histogram.
Theoretical predictions for the hadroproduction of more exclusive final states,
such as $l^+l^-+j$ \cite{Denner:2011vu} or $\nu\bar{\nu}+j$
\cite{Denner:2012ts}, require a different mode of experimental data analysis.
In this sense, the results presented here are not already included in
Refs.~\cite{Denner:2011vu,Denner:2012ts}, the more so as QCD-electroweak
interference contributions of the type mentioned above were neglected there.
Our detailed study confirmed the expectation \cite{Denner:2011vu,Denner:2012ts}
based on the analysis of $l^+\nu+j$ hadroproduction \cite{Denner:2009gj} that
those interference contributions are numerically small, for the small-$p_T$
regime.

\section*{Acknowledgments}

We thank Maria Vittoria Garzelli for a numerical cross check of our NLO QCD
calculation.
B.A.K. is indebted to the Max Planck Institut f\"ur Physik for the hospitality
during a visit, when part of his work on this manuscript was performed.
This work was supported in part by the German Federal Ministry for Education
and Research BMBF through Grant No.~05H12GUE.

%!!!!!!!!!!!!!!!!!!!!!!!!!!!!!!!!!!!
\appendix
%%%%%%%%%%%%%%%%%%%%%%%%%%%%%%%%%%%%%%%%%%%%%%%%%%%%%%%%%%%%

\boldmath
\section{Cross section of $\bar{q}+q\to Z+X$ through
$\mathcal{O}(\alpha^2\alpha_s)$}
\label{sec:a}
\unboldmath

In this appendix, we consider $\mathcal{O}(\alpha^2\alpha_s)$ contributions to
$\bar q+q\to Z+X$.
This includes the virtual QED corrections to subprocess (\ref{qQ_Zg}), the
virtual QCD corrections to subprocess (\ref{qQ_Zy}), the real corrections from
subprocess (\ref{qQ_Zgy}), and the interference contributions from subprocess
(\ref{qQ_ZqQ}) involving a virtual photon or $Z$ boson.
We have
\begin{eqnarray}
  \frac{s\,d\sigma^{\bar{q}q}}{dt\,du} &=& 
    \frac{2\pi\alpha^2 Q_q^2(v_q^2+a_q^2)}{N_c s} 
    \left[ \delta(s_2) A_0
\vphantom{\frac{v_q^4+6v_q^2a_q^2+a_q^4}{v_q^2+a_q^2}}\right.
\nonumber\\
&&{}+\left.
C_F\frac{\alpha_s}{\pi} \left( \delta(s_2) A_1 + A_2 
               + A_3 + \frac{v_q^4+6v_q^2a_q^2+a_q^4}{v_q^2+a_q^2} A_4\right)
\right] \,,
\label{eq:qQ}
\end{eqnarray}
where
\begin{align}
A_0 &= \frac{t}{u} + \frac{u}{t} + 2\left( 
    \frac{s}{t} + \frac{s}{u} + \frac{s^2}{tu}\right),
\\
%%%%
%% delta-part for qbarq
A_1 &= 
       A_0 \Bigg[ 
         L_{\mu_F} ( 2 L_t + 2 L_u - 4 L_A - 3 )
       + 4 L_A ( L_A + L_s - L_t - L_u - 1 )
       + (L_t + L_u)^2
         \Bigg]
\nonumber\\
&{}  
       + L_s^2 \Bigg( - \frac{2s^2}{tu} - A_0 \Bigg)
       + L_s \Bigg( \frac{8s}{t+u} + \frac{4s^2}{(t+u)^2} \Bigg)
\nonumber\\
&{}  
       + \Bigg( 2A_0 + \frac{4s^2}{tu} \Bigg)
         \Bigg( L_s \ln\frac{s-Q^2}{Q^2} 
            - {\rm Li}_2\Big( \frac{Q^2}{s}\Big) \Bigg)
       + \zeta(2) \Bigg( \frac{4s^2}{tu} + 8A_0 \Bigg)
\nonumber\\
&{}
      + \Bigg\{ L_t \Bigg( \frac{t+4s}{s+u} + \frac{st}{(s+u)^2} \Bigg)
%       + L_u \Bigg( \frac{u+4s}{s+t} + \frac{su}{(s+t)^2} \Bigg)
       + L_s L_t \Bigg( \frac{2(2s+t)}{u} - 4A_0 \Bigg)
\nonumber\\
%     + L_s L_u \Bigg( \frac{2(2s+u)}{t} - 4A_0 \Bigg)
&{}  
       + \frac{2(2s^2 + 2su + u^2)}{tu} 
         \Bigg( L_t \ln\frac{s+u}{Q^2} 
            + {\rm Li}_2\Big( \frac{t}{Q^2}\Big) \Bigg)
%\nonumber\\
%&  
%       + \frac{2(2s^2 + 2st + t^2)}{tu} 
%         \Bigg( L_u \ln\Big(\frac{s+t}{Q^2}\Big) 
%            + {\rm Li}_2\Big( \frac{u}{Q^2}\Big) \Bigg)
\nonumber\\
&{}  
%        - \frac{9u+17s}{t} 
				- \frac{9t+17s}{u}
%        - \frac{16s^2}{tu}
        - \frac{8s^2}{tu}
%          + \frac{4s}{t+u}
          + \frac{2s}{t+u}
%          + \frac{s}{s+u}
          + \frac{s}{s+t}\Biggr\} + \Bigg\{ u \leftrightarrow t \Bigg\} \,,
% NODELTA part for qbarq
\\
A_2 &= 
       \Bigg[ 4A_0 \Bigg( 
          - \ln\frac{(s_2-t)(s_2-u)}{s^2}
          - L_{\mu_F}
          + 2\ln\frac{s_2}{Q^2} - L_s
        \Bigg) \frac{1}{s_2} \Bigg]_{A+}
\nonumber\\
&
%*** ftu
     + \ln\frac{tu-Q^2 s_2}{(s_2-t)(s_2-u)}
      \Bigg(
         \frac{4s^2(s-Q^2)}{tu(s_2-t)(s_2-u)}
        + \frac{2(2s+u-t)}{t(s_2-t)}
        + \frac{2(2s+t-u)}{u(s_2-u)}
\nonumber\\
&
\qquad        + \frac{4 Q^2 A_0 -2u-2t-2Q^2-6s}{tu-Q^2 s_2}
       + \frac{2(t+u+s+Q^2)}{tu}
       \Bigg)
\nonumber\\
&
%** log
     + \ln\frac{(s_2-t)(s_2-u)}{s s_2}
       \Bigg(
         \frac{4s^2(s-Q^2)}{tu(s_2-t)(s_2-u)}
        + \frac{2(2s+u-t)}{t(s_2-t)}
        + \frac{2(2s+t-u)}{u(s_2-u)}
      \Bigg)
\nonumber\\
&
%** log
      + \ln\frac{\mu_F^2}{s_2}
        \Bigg(
         \frac{s}{(s_2-t)^2}
       + \frac{s}{(s_2-u)^2}
       - \frac{2s+u+t}{t(s_2-t)}
       - \frac{2s+t+u}{u(s_2-u)}
       + \frac{Q^2}{t^2}
       + \frac{Q^2}{u^2}
\nonumber\\
&
\qquad       - \frac{4 Q^2 A_0 -2u-2t-2Q^2-6s}{tu-Q^2 s_2}
      \Bigg)
\nonumber\\
&
%** no-log
       + \frac{s}{(s_2-t)^2}
       + \frac{s}{(s_2-u)^2}
       + \frac{2(u/t+s/t-s/u)}{s_2-t}
       + \frac{2(t/u+s/u-s/t)}{s_2-u}
\nonumber\\
&
       + \frac{2(s - Q^2 + Q^2 u/t + Q^2 t/u)}{tu-Q^2 s_2}
       + \frac{2(t+u-s)}{tu}\,,
\\
%%%%
A_3 &=
    \frac{1}{\lambda}
%      \ln\Big(\frac{s+Q^2-s_2+\lambda}{s+Q^2-s_2-\lambda}\Big)
      L_\lambda
      \Bigg(
        \frac{3 s (t-u)^2 (t+u) (4 Q^2 s + (t+u)^2)}{2 t u\lambda^4}
\nonumber\\
&
\qquad
       - \frac{2 s (t^3+u^3)+Q^2 (-4 s (t-u)^2+2 s^2 (t+u)-(t-u)^2 (t+u))}
              {t u\lambda^2}
\nonumber\\
&
\qquad
       + \frac{(t+u) (4 s^2+t^2+u^2+2 s (t+u))}{s_2 t u}
       + \frac{6(s-Q^2)(t^2+u^2)}{stu}
       + \frac{2(6 s+t+u)}{s+Q^2-s2}
\nonumber\\
&
\qquad
       + \frac{(9 s-18 Q^2+8 Q^4/s) (t+u)}{2 t u}
       + \frac{3(t^3 + t^2 u + t u^2 + u^3)}{s t u}
       - \frac{4 Q^2}{s}
       + 2
      \Bigg)
\nonumber\\
&
       + \frac{1}{2st} 
%         \ln\Big(\frac{s Q^2(s_2-t)^2}{(s_2(2Q^2-u)-Q^2 t)^2}\Big) 
         L_{\lambda t}
        \Bigg(
            \frac{4 s^2 + t^2 + 4 s u + u^2}{s+Q^2-s_2}
             +2 Q^2-4 s-t-3 u \Bigg)
\nonumber\\
&
       + \frac{1}{2su} 
%         \ln\Big(\frac{s Q^2(s_2-u)^2}{(s_2(2Q^2-t)-Q^2 u)^2}\Big) 
         L_{\lambda u}
        \Bigg(
            \frac{4 s^2 + u^2 + 4 s t + t^2}{s+Q^2-s_2}
             +2 Q^2-4 s-u-3 t \Bigg)
\nonumber\\
&
       + \frac{1}{2t} L_{st}
          \Bigg(
             \frac{-2 s_2+3 t+u}{s}
           + \frac{2(2 s^2+2 s u+u^2)}{s_2 u}
\nonumber\\
&
\qquad
           - \frac{2(2 s-2 s_2+t+2 u)}{t}
           + \frac{4 s^2+t^2+4 s u+u^2}{s(s+Q^2-s_2)}
          \Bigg)
\nonumber\\
&
       + \frac{1}{2u}L_{su}
          \Bigg(
             \frac{-2 s_2+3 u+t}{s}
           + \frac{2(2 s^2+2 s t+t^2)}{s_2 t}
\nonumber\\
&
\qquad
           - \frac{2(2 s-2 s_2+u+2 t)}{u}
           + \frac{4 s^2+u^2+4 s t+t^2}{s(s+Q^2-s_2)}
          \Bigg)
\nonumber\\
&
%%% no-logs
     + \frac{3(t - u)^2 (t + u) (-2Q^2(2s+t+u)+(t+u) (4 s + t + u))}
             {2tu\lambda^4}
\nonumber\\
&
     - \frac{t^3 + t^2 u + t u^2 + u^3 - Q^2 (t + u) (2 s + t + u) + 
                    s (6 t^2 - 4 t u + 6 u^2)}{tu\lambda^2}
\nonumber\\
&
       + \frac{s^2}{u(s_2-t)^2}
       + \frac{s^2}{t(s_2-u)^2}
       + \Bigg(2 + \frac{3s^2}{tu}\Bigg) 
          \Bigg(\frac{1}{s_2-t} + \frac{1}{s_2-u}\Bigg)
\nonumber\\
&
       + \frac{1}{s}\Bigg(\frac{1}{t^2}+\frac{1}{u^2}\Bigg)
           \Big(4Q^4 + 2tu - Q^2(7s + 4(t + u)) \Big)
       + \frac{3s(t + u)-4tu}{2stu} \,,
\label{eq:a3}\\
A_4 &=
        \frac{L_\lambda}{\lambda^5}  
         \frac{3s(t-u)^2(t+u)}{4tu} \Bigg(
      \frac{s(t+u)^2(s+s_2-Q^2)}{(s-Q^2)(s_2-Q^2)}
     +4sQ^2+(t+u)^2 \Bigg)
\nonumber\\
&
       + \frac{L_\lambda}{\lambda^3} \Bigg(
          \frac{s (-3 t^3 + t^2 u + t u^2 - 3 u^3) + Q^2(4 s (t - u)^2 - 2 s^2 (t + u) + (t - u)^2 (t + u))}{2 t u}
\nonumber\\
&
         - \frac{s^2 (t^3 + t^2 u + t u^2 + u^3)}{2 (s-Q^2) t u}
         + \frac{s ((t^2 - u^2)^2 - 2 s (t^3 + u^3))}{2 t u (s_2-Q^2)} \Bigg)
\nonumber\\
&
       + \frac{L_\lambda}{\lambda} \Bigg(
          - 2
          + 2 \frac{s_2^2}{u (s-Q^2)}
          + \frac{s_2^2}{u (s_2-Q^2)}
          + 2 \frac{s_2^2}{t (s-Q^2)}
          + \frac{s_2^2}{t (s_2-Q^2)}
\nonumber\\
&
          + 2 \frac{s_2^2}{s u}
          + 2 \frac{s_2^2}{s t}
          - \frac{u s_2}{t (s-Q^2)}
          - \frac{u s_2}{t (s_2-Q^2)}
          - 4 \frac{u s_2}{s t}
          + \frac{1}{2} \frac{u^2}{t (s-Q^2)}
          + \frac{1}{2} \frac{u^2}{t (s_2-Q^2)}
\nonumber\\
&
          + \frac{7}{2} \frac{u^2}{t s_2}
          - \frac{u^2}{s (s+Q^2-s_2)}
          + \frac{7}{2} \frac{u^2}{s t}
          - \frac{t s_2}{u (s-Q^2)}
          - \frac{t s_2}{u (s_2-Q^2)}
          - 4 \frac{t s_2}{s u}
\nonumber\\
&
          - 2 \frac{t u}{s (s+Q^2-s_2)}
          + \frac{1}{2} \frac{t^2}{u (s-Q^2)}
          + \frac{1}{2} \frac{t^2}{u (s_2-Q^2)}
          + \frac{7}{2} \frac{t^2}{u s_2}
          - \frac{t^2}{s (s+Q^2-s_2)}
          + \frac{7}{2} \frac{t^2}{s u}
\nonumber\\
&
          - \frac{1}{2} \frac{s s_2}{u (s-Q^2)}
          - \frac{s s_2}{u (s_2-Q^2)}
          - \frac{1}{2} \frac{s s_2}{t (s-Q^2)}
          - \frac{s s_2}{t (s_2-Q^2)}
          + \frac{s u}{(s+Q^2-s_2) (s-Q^2)}
\nonumber\\
&
          + \frac{1}{2} \frac{s u}{t (s-Q^2)}
          + \frac{1}{2} \frac{s u}{t (s_2-Q^2)}
          + 10 \frac{s u}{t s_2}
          + \frac{s t}{(s+Q^2-s_2) (s-Q^2)}
          + \frac{1}{2} \frac{s t}{u (s-Q^2)}
\nonumber\\
&
          + \frac{1}{2} \frac{s t}{u (s_2-Q^2)}
          + 10 \frac{s t}{u s_2}
          + 6 \frac{s^2}{(s+Q^2-s_2) (s-Q^2)}
          + \frac{3}{4} \frac{s^2}{u (s-Q^2)}
          + \frac{9}{4} \frac{s^2}{u (s_2-Q^2)}
\nonumber\\
&
          + 8 \frac{s^2}{u s_2}
          + \frac{3}{4} \frac{s^2}{t (s-Q^2)}
          + \frac{9}{4} \frac{s^2}{t (s_2-Q^2)}
          + 8 \frac{s^2}{t s_2}
          - \frac{17}{4} \frac{s}{t}
          - \frac{17}{4} \frac{s}{u}
          + 20 \frac{s}{s_2}
\nonumber\\
&
          + 6 \frac{s}{(s+Q^2-s_2)}
          + 3 \frac{s}{(s_2-Q^2)}
          + \frac{9}{2} \frac{t}{s}
          + \frac{3}{2} \frac{t}{u}
          + \frac{19}{2} \frac{t}{s_2}
          + 3 \frac{t}{(s+Q^2-s_2)}
          + \frac{1}{2} \frac{t}{(s_2-Q^2)}
\nonumber\\
&
          + \frac{5}{2} \frac{t}{(s-Q^2)}
          + \frac{9}{2} \frac{u}{s}
          + \frac{3}{2} \frac{u}{t}
          + \frac{19}{2} \frac{u}{s_2}
          + 3 \frac{u}{(s+Q^2-s_2)}
          + \frac{1}{2} \frac{u}{(s_2-Q^2)}
          + \frac{5}{2} \frac{u}{(s-Q^2)}
\nonumber\\
&
          - 6 \frac{s_2}{s}
          - \frac{3}{2} \frac{s_2}{t}
          - \frac{3}{2} \frac{s_2}{u}
          - 2 \frac{s_2}{(s_2-Q^2)}
          - 6 \frac{s_2}{(s-Q^2)}
          \Bigg)
\nonumber\\
&
       + \frac{1}{\lambda^4} \frac{3 (t - u)^2 (t + u)}{4tu(Q^2-s) (s_2-Q^2} 
         \Bigg(
              4 Q^6 (2 s + t + u) - 4 Q^4 \Big(3 s^2 + 6 s (t + u) + (t + u)^2\Big)
\nonumber\\
&
             + Q^2 \Big( 4 s^3 + 28 s^2 (t + u) + 14 s (t + u)^2 + (t + u)^3 \Big) 
               - s (t + u) \Big(8 s^2 + 9 s (t + u) 
\nonumber\\
&
               + 2 (t + u)^2\Big) \Bigg)
       + \frac{1}{\lambda^2}  \frac{1}{4 tu(s-Q^2)(s_2-Q^2)}
               \Bigg(
                  4 Q^6 (t + u) (2 s + t + u) 
\nonumber\\
&
        - 4 Q^4 \Big(3 s^2 (t + u) + (t + u)^3 + s (13 t^2 - 2 t u + 13 u^2)\Big) 
        +  Q^2 \Big( 4 s^3 (t + u) 
\nonumber\\
&
                     + (t + u)^4  + 8 s^2 (9 t^2 - 4 t u + 9 u^2) + 
                  8 s (4 t^3 + 3 t^2 u + 3 t u^2 + 4 u^3) \Big)  
\nonumber\\
&
                   - s \Big(8 s^2 (3 t^2 - 2 t u + 3 u^2) + (t + u)^2 (5 t^2 - 2 t u + 5 u^2)
\nonumber\\
&
                   + 12 s (2 t^3 + t^2 u + t u^2 + 2 u^3) \Big)  \Bigg)
\nonumber\\
&
       + 2 \frac{s_2^2}{u^2 (s-Q^2)}
          + \frac{3}{2} \frac{s_2^2}{u^2 (s_2-Q^2)}
          + 2 \frac{s_2^2}{t^2 (s-Q^2)}
\nonumber\\
&
          + \frac{3}{2} \frac{s_2^2}{t^2 (s_2-Q^2)}
          - 2 \frac{u s_2}{t^2 (s-Q^2)}
          - \frac{3}{2} \frac{u s_2}{t^2 (s_2-Q^2)}
          - \frac{u^2}{t^2 (s_2-t)}
          - 2 \frac{t s_2}{u^2 (s-Q^2)}
\nonumber\\
&
          - \frac{3}{2} \frac{t s_2}{u^2 (s_2-Q^2)}
          - \frac{t^2}{u^2 (s_2-u)}
          - 2 \frac{s s_2}{u^2 (s-Q^2)}
          - \frac{3}{2} \frac{s s_2}{u^2 (s_2-Q^2)}
          - 2 \frac{s s_2}{t^2 (s-Q^2)}
\nonumber\\
&
          - \frac{3}{2} \frac{s s_2}{t^2 (s_2-Q^2)}
          - 2 \frac{s u}{t^2 (s_2-t)}
          - 2 \frac{s t}{u^2 (s_2-u)}
          + 3 \frac{s^2}{(s_2-u)^3}
          + 3 \frac{s^2}{(s_2-t)^3}
\nonumber\\
&
          + \frac{s^2}{u (s_2-u)^2}
          + \frac{3}{2} \frac{s^2}{u (s_2-t) (s_2-Q^2)}
          - \frac{13}{2} \frac{s^2}{u (s_2-t)^2}
          - \frac{s^2}{u^2 (s_2-u)}
\nonumber\\
&
          + \frac{3}{2} \frac{s^2}{t (s_2-u) (s_2-Q^2)}
          - \frac{13}{2} \frac{s^2}{t (s_2-u)^2}
          + \frac{s^2}{t (s_2-t)^2}
          + 3 \frac{s^2}{t u (s_2-Q^2)}
\nonumber\\
&
          + \frac{1}{2} \frac{s^2}{t u (s_2-u)}
          + \frac{1}{2} \frac{s^2}{t u (s_2-t)}
          - \frac{s^2}{t^2 (s_2-t)}
          + \frac{1}{2} \frac{s^2 s_2}{u (s_2-t)^2 (s_2-Q^2)}
\nonumber\\
&
          + \frac{1}{2} \frac{s^2 s_2}{t (s_2-u)^2 (s_2-Q^2)}
          + 3 \frac{s^3}{u (s_2-t)^3}
          + 3 \frac{s^3}{t (s_2-u)^3}
          + \frac{s^3}{t u (s_2-u)^2}
\nonumber\\
&
          + \frac{s^3}{t u (s_2-t)^2}
          + \frac{s}{(s_2-u) (s-Q^2)}
          - 4 \frac{s}{(s_2-u)^2}
          + \frac{s}{(s_2-t) (s-Q^2)}
\nonumber\\
&
          - 4 \frac{s}{(s_2-t)^2}
          + \frac{1}{4} \frac{s}{u (s-Q^2)}
          - \frac{1}{4} \frac{s}{u (s_2-Q^2)}
          + 5 \frac{s}{u (s_2-t)}
          + \frac{1}{2} \frac{s}{u^2}
\nonumber\\
&
          + \frac{1}{4} \frac{s}{t (s-Q^2)}
          - \frac{1}{4} \frac{s}{t (s_2-Q^2)}
          + 5 \frac{s}{t (s_2-u)}
          + \frac{1}{2} \frac{s}{t^2}
          + \frac{t}{u (s-Q^2)}
          + \frac{1}{4} \frac{t}{u (s_2-Q^2)}
\nonumber\\
&
          + \frac{1}{2} \frac{t}{u^2}
          - 2 \frac{t}{s^2}
          + \frac{u}{t (s-Q^2)}
          + \frac{1}{4} \frac{u}{t (s_2-Q^2)}
          + \frac{1}{2} \frac{u}{t^2}
          - 2 \frac{u}{s^2}
          - 2 \frac{s_2}{u (s-Q^2)}
\nonumber\\
&
          - \frac{s_2}{u (s_2-Q^2)}
          + \frac{1}{2} \frac{s_2}{u^2}
          - 2 \frac{s_2}{t (s-Q^2)}
          - \frac{s_2}{t (s_2-Q^2)}
          + \frac{1}{2} \frac{s_2}{t^2}
          + 2 \frac{s_2}{s^2}
          - 5 \frac{1}{s}
          - \frac{3}{4} \frac{1}{t}
\nonumber\\
&
          - \frac{3}{4} \frac{1}{u}
          + 2 \frac{1}{(s_2-t)}
          + 2 \frac{1}{(s_2-u)}
          - \frac{1}{2} \frac{1}{(s_2-Q^2)}
          - \frac{1}{(s-Q^2)}
\nonumber\\
&
       + \Bigg\{
          L_t   \Bigg(
           \frac{t^2}{u^2 (s_2-u)}
          - \frac{t^3}{u^2 (s_2-u)^2}
          - 4 \frac{s t}{(s_2-u)^3}
          + 2 \frac{s t}{u^2 (s_2-u)}
          - 3 \frac{s t^2}{u^2 (s_2-u)^2}
\nonumber\\
&
          - 12 \frac{s^2}{(s_2-u)^3}
          + \frac{s^2}{u^2 (s_2-u)}
          + 8 \frac{s^2}{t (s_2-u)^2}
          - 4 \frac{s^2}{t u (s_2-u)}
          + 4 \frac{s^2}{t u s_2}
          + 3 \frac{s^2 t}{(s_2-u)^4}
\nonumber\\
&
          + \frac{s^2 t}{u (s_2-u)^3}
          - 3 \frac{s^2 t}{u^2 (s_2-u)^2}
          + 6 \frac{s^3}{(s_2-u)^4}
          + 2 \frac{s^3}{u (s_2-u)^3}
          - \frac{s^3}{u^2 (s_2-u)^2}
\nonumber\\
&
          - 8 \frac{s^3}{t (s_2-u)^3}
          + 3 \frac{s^4}{t (s_2-u)^4}
          + \frac{s^4}{t u (s_2-u)^3}
          + 9 \frac{s}{(s_2-u)^2}
          - 4 \frac{s}{u (s_2-u)}
          + 6 \frac{s}{u s_2}
\nonumber\\
&
          - 4 \frac{s}{t (s_2-u)}
          + 4 \frac{s}{t s_2}
          + 2 \frac{t}{(s_2-u)^2}
          - \frac{t}{u (s_2-u)}
          + \frac{5}{2} \frac{t}{u s_2}
          + \frac{3}{2} \frac{t}{s u}
          + \frac{u}{t s_2}
          - \frac{s_2}{s u}
\nonumber\\
&
          + \frac{3}{2} \frac{1}{s}
          + \frac{1}{u}
          + 3 \frac{1}{s_2}
          - 3 \frac{1}{(s_2-u)}
          \Bigg)
       + L_{st} \Bigg(
               \frac{2 Q^2 + t - u}{4 t(s-Q^2)} 
             + \frac{4 s^2 + t^2 + 4 s u + u^2}{4t (s-Q^2) (Q^2+ s - s_2)} 
\nonumber\\
&
             + \frac{2st(s+u) - 2s_2 u Q^2 + tu^2}{2 t^2 u (s_2-Q^2)}
              \Bigg)
       + \frac{2 s_2^2 - 2 s_2 t + t^2}{2t(s-Q^2)(Q^2 +s-s_2)} L_{\lambda t}
\nonumber\\
&
       + I_4(1,1,t,-1)   \Bigg(
           \frac{4 s^3 - u (t + u)^2 + 4 s^2 (t + 2 u) + s (t^2 + 2 t u + 3 u^2)}{4s(Q^2+s-s_2)} 
\nonumber\\
&
          + \frac{15}{2} s_2
          - \frac{3}{4} u
          + \frac{13}{4} t
          + \frac{35}{2} s
          - 6 \frac{s_2^2}{s}
          - 6 \frac{s_2^2}{t}
          + \frac{s_2^2}{u}
          + 2 \frac{s_2^3}{s t}
          + \frac{23}{2} \frac{u s_2}{s}
          + 10 \frac{u s_2}{t}
\nonumber\\
&
          - 6 \frac{u s_2^2}{s t}
          - \frac{23}{4} \frac{u^2}{s}
          - 2 \frac{u^2}{t}
          - 8 \frac{u^2}{s_2}
          + \frac{13}{2} \frac{u^2 s_2}{s t}
          - \frac{5}{2} \frac{u^3}{t s_2}
          - \frac{5}{2} \frac{u^3}{s t}
          + 6 \frac{t s_2}{s}
          - 3 \frac{t s_2}{u}
\nonumber\\
&
          - \frac{21}{4} \frac{t u}{s}
          - \frac{19}{2} \frac{t u}{s_2}
          - 2 \frac{t^2}{s}
          + 3 \frac{t^2}{u}
          - 5 \frac{t^2}{s_2}
          - \frac{t^3}{u s_2}
          + 2 \frac{s s_2}{t}
          - 6 \frac{s s_2}{u}
          + 6 \frac{s u}{t}
\nonumber\\
&
          - 24 \frac{s u}{s_2}
          - \frac{17}{2} \frac{s u^2}{t s_2}
          + 12 \frac{s t}{u}
          - 21 \frac{s t}{s_2}
          - 6 \frac{s t^2}{u s_2}
          + 6 \frac{s^2}{t}
          + 12 \frac{s^2}{u}
          - 24 \frac{s^2}{s_2}
          - 10 \frac{s^2 u}{t s_2}
\nonumber\\
&
          - 12 \frac{s^2 t}{u s_2}
          - 8 \frac{s^3}{u s_2}
          - 4 \frac{s^3}{t s_2}
          \Bigg)
\nonumber\\
&
       + I_4(1,1,t,1)   \Bigg(
              - \frac{s_2 (s + u)}{2s} 
              - \frac{(s + u) (4 s^2 - t^2 + t u + 4 u^2 + 2 s (t + 4 u))}{4st} 
\nonumber\\
&
              - \frac{4 s^3 - u (t + u)^2 + 4 s^2 (t + 2 u) + s (t^2 + 2 t u + 3 u^2)}{4s(Q^2+s-s_2)}
                 \Bigg)
\nonumber\\
&
       + \frac{2 Q^2(s+Q^2)^2}{s} I_4(1,2,t,-1)
        \Bigg\}
        + \Bigg\{ t \longleftrightarrow u  \Bigg\}\,. 
\end{align}
Here, $A=s_2^{\rm max}$ as defined in Eq.~(\ref{eq:alt}),
$\lambda=\lambda(s,Q^2,s_2)$ with\break
$\lambda(x,y,z)=\sqrt{x^2+y^2+z^2-2(xy+yz+zx)}$ being K\"all\'en's function,
we have introduced the short-hand notations
\begin{eqnarray}
L_s &=& \ln\frac{s}{Q^2},  \quad 
L_t = \ln\frac{-t}{Q^2}, \quad
L_u = \ln\frac{-u}{Q^2}, \quad
L_A = \ln\frac{A}{Q^2}, \quad
L_{\mu_F} = \ln\frac{\mu_F^2}{Q^2},
\nonumber\\
L_{st} &=& \ln\frac{st^2}{Q^2(s_2-t)^2}, \quad
L_{su} = \ln\frac{su^2}{Q^2(s_2-u)^2}, \quad
L_\lambda =  \ln\frac{s+Q^2-s_2+\lambda}{s+Q^2-s_2-\lambda}, 
\nonumber\\
L_{\lambda t} &=& \ln\frac{s Q^2(s_2-t)^2}{[s_2(2Q^2-u)-Q^2 t]^2},\quad
L_{\lambda u} = \ln\frac{s Q^2(s_2-u)^2}{[s_2(2Q^2-t)-Q^2 u]^2},
\end{eqnarray}
and we have adopted the following phase-space integrals from Appendix~C in
Ref.~\cite{Beenakker:1988bq}:
\begin{align}
I_4(1,1,t,1) &= 
  \frac{1}{\sqrt{X_+}}
	\ln\frac{ u (Q^2-s_2) + 2 Q^2 s + \sqrt{X_+}}{u (Q^2-s_2) + 2 Q^2 s  - \sqrt{X_+}} \,,
%  xtp = u*(QQ-s2) + 2*QQ*s;
\nonumber\\	
I_4(1,1,t,-1) &=  
  \frac{1}{\sqrt{X_-}}
	\ln \frac{ 2 Q^4- Q^2 (t+u) - s t + \sqrt{X_-}}{ 2 Q^4 -Q^2 (t+u) - s t  - \sqrt{X_-}} \,,
\nonumber\\
%  xtm = 2*SQR(QQ) - QQ*(t+u) - s*t;
I_4(1,2,t,-1) &=
    \frac{4 Q^6 + st(t + u) - 2 Q^4(s + 2(t + u)) + 
       Q^2(-2 s^2 - 2s(t + u) + (t + u)^2)}{2 Q^2 s X_-} 
\nonumber\\
   & + \frac{2 Q^6 - s (s+u) t + Q^2 u (s+t+u) - Q^4 (2s+3u+t)}{2 X_-}I_4(1,1,t,-1) \,,
\label{eq:psintegrals}
\end{align}
where
\begin{eqnarray}
%  XXtp = sqrt(  SQR(xtp) - 4*QQ*QQ*s*(s+u)  );
%  XXtm = sqrt(  SQR(xtm) - 4*QQ*QQ*s*(s+u)  );
X_+ &=& [u (Q^2-s_2)+2 Q^2 s ]^2 - 4 Q^4 s (s+u)  ,
\nonumber\\
X_- &=& [2 Q^4- Q^2 (t+u) - s t ]^2 - 4 Q^4 s (s+u) .
\end{eqnarray}
%Notice that the three terms proportional to $1/s_2$ in the expression for $A_3$
%in Eq.~(\ref{eq:a3}) conspire to yield a finite result in the limit $s_2\to0$
%of vanishing invariant mass of the $q\bar{q}$ pair in the final state.

%%%%%%%%%%%%%%%%%%%%%%%%%%%555
\boldmath
\section{Cross section of $q+g\to Z+X$ through $\mathcal{O}(\alpha^2\alpha_s)$}
\label{sec:b}
\unboldmath

In this appendix, we consider $\mathcal{O}(\alpha^2\alpha_s)$ contributions to
$q+g\to Z+X$.
This includes the virtual QED corrections to subprocess (\ref{qg_Zq}) and the
real corrections from subprocess (\ref{qg_Zqy}).
We have
\begin{equation}
  \frac{s\,d\sigma^{qg}}{dt\,du} = 
    \frac{2\pi\alpha\alpha_s Q_q^2(v_q^2+a_q^2)C_F}{(N_c^2-1)s}
    \left[ \delta(s_2) B_0 + 
      \frac{\alpha}{\pi} \left( \delta(s_2) B_1 + B_2 \right) \right]\,,
\label{eq:qg}
\end{equation}
where
\begin{eqnarray}
B_0 &=& - \left[ \frac{s}{t} + \frac{t}{s} + 2\left( 
    \frac{u}{t} + \frac{u}{s} + \frac{u^2}{st}\right) \right],
\end{eqnarray}
\begin{align}
%%%%
%% delta-part for qg
 B_1 &= 
%*** delta-terms
     B_0 \Bigg[
             L_{\mu_F} ( L_u - L_A - \frac{3}{4} )
          + \frac{1}{2} L_A^2 - \frac{3}{4} L_A 
        \Bigg]
\nonumber\\
&
%     + \frac{t^2+2 t u+2 u^2}{2 s t}
     + \frac{(t+u)^2+ u^2}{2 s t}
        \Bigg( 2 {\rm Li}_2\Big(\frac{Q^2}{s}\Big) + L_s^2
          + 2 L_s L_u - 2 L_s \ln\frac{s-Q^2}{Q^2} \Bigg)
\nonumber\\
&
%         - \frac{s^2+2 s u+2 u^2}{s t}
         - \frac{(s+u)^2 + u^2}{s t}
          \Bigg( {\rm Li}_2\Big(\frac{t}{Q^2}\Big) 
           - L_t L_u + L_t \ln\frac{s+u}{Q^2}  \Bigg)
\nonumber\\
&
%         - \frac{s^2+t^2+2 s u+2 t u+4 u^2}{s t}
         - \frac{(s+u)^2+(t+u)^2+2 u^2}{s t}
             \Bigg( {\rm Li}_2\Big(\frac{u}{Q^2}\Big) 
             + L_u \ln\frac{s+t}{Q^2} \Bigg)
\nonumber\\
&
%         + \frac{s^2 + t^2 + 2 s u + 2 t u + 2 u^2}{2 s t} L_u^2
         + \frac{(s+u)^2 + (t+u)^2}{2 s t} L_u^2
         - \frac{2u(2s+2t+u)}{(s+t)^2} L_u
         + \Bigg( - \frac{2u+t}{s+u} + \frac{s t}{2(s+u)^2} \Bigg) L_t
\nonumber\\
&
         - \Bigg( \frac{s+4 u}{2(t+u)} + \frac{s u}{2(t+u)^2} \Bigg) L_s
         - \frac{2(2 s^2 + 4 s u + 5 u^2}{s t} \zeta(2)
         - \frac{u}{2(t+u)}
\nonumber\\
&
          - \frac{2 u}{s+t}
          + \frac{s}{2(s+u)}
         + \frac{11( s^2+t^2) - 2 s t  + 20( s u + t u) + 18 u^2}{4 s t}\,,
\end{align}
\begin{align}
%%
%%% terms with no delta(s2)!!!
%%
B_2 &=
      \frac{1}{\lambda}
%         \ln\Big(\frac{s+Q^2-s_2+\lambda}{s+Q^2-s_2-\lambda}\Big)
         L_\lambda
         \Bigg[
          \frac{3 s (t - u)^2 (t + u) (-2 Q^2 + t + u)}{8 t\lambda^4}
\nonumber\\
&
\qquad
       +\frac{1}{\lambda^2}\Bigg(
             \frac{(t-u)^2 (t+u)-s (t^2+u^2)}{4t}
            + \frac{(t-u)(t + u)^2}{8s}
\nonumber\\
&
\qquad
            + \frac{Q^2(-t^3+s^2(3t-u)+tu^2-2s(3t^2-4tu+u^2))}{4st}
           \Bigg)
\nonumber\\
&
\qquad
        + \frac{64Q^4+7s^2+7t^2+21tu+16u^2+2s(t+u)-2Q^2(30s+15t+16u)}{8st}
\nonumber\\
&
\qquad
        - \frac{t^3 + 3 t^2 u + 4 t u^2 + 2 u^3}{2st(s_2)_{A+}}
         \Bigg]
      + L_{\mu_F} \frac{s^2 + t^2 + 2 s u + 2 t u + 2  u^2}{s t(s_2)_{A+}}
\nonumber\\
&
%      + \ln\Big(\frac{su^2}{Q^2(s_2-u)^2}\Big) \Bigg[
      + L_{su} \Bigg[
         - \frac{t^2 + 2 t u + 2 u^2}{2 s t (s_2)_{A+}}
         + \frac{1}{2 s} - \frac{1}{2 t} + \frac{Q^2}{s t} 
         + \frac{Q^2(Q^2-t-s)}{2 s u^2} 
\nonumber\\
&
\qquad
         - \frac{1}{u} + \frac{Q^2}{s u} - \frac{t}{2 s u} 
         + \frac{u}{s t}
         - \frac{2 Q^4 - 2 Q^2 (s + t) + (s + t)^2}{2 s t (Q^2-u)}
           \Bigg]   
\nonumber\\
&
       + \ln\frac{tu-s_2Q^2}{(s_2-t)(s_2-u)} \Bigg[
        - \frac{s^2 + 2 s u + 2 u^2}{s t (s_2)_{A+}}
        + \Bigg( Q^2 + t 
				- \frac{2 Q^2 }{st}\Big((s-u)^2+(t-u)^2\Big)  
%        + \Bigg( s^2 + t - \frac{2 Q^2 s}{t}  
%            - \frac{2 Q^2 t}{s} 
\nonumber\\
&
\qquad
%        + \frac{2 t^2}{s} + \frac{4 Q^2 u}{s} 
%           - \frac{4 Q^2 u}{t} - \frac{4 t u}{s} + \frac{4 u^2}{s} 
%           - \frac{4 Q^2 u^2}{s t} 
        + \frac{t^2+(t-2u)^2}{s} 
           \Bigg)\frac{1}{tu-s_2 Q^2}
%\nonumber\\
%&
%\qquad
%        + \frac{2 Q^4 - 2 Q^2 (s + t) + (s + t)^2}{st(Q^2-u)}
        + \frac{2 Q^2(u-s_2) + (s + t)^2}{st(Q^2-u)}
        + \frac{-2 Q^2 + 3 t - 2 u}{s t}
          \Bigg]
\nonumber\\
&
%       + \ln\Big(\frac{sQ^2(s_2-t)^2}{[s_2(2Q^2-t)-Q^2u]^2}\Big) 
       + L_{\lambda t}
        \Bigg(
%           \frac{2 Q^4 - 2 Q^2 (s + t) + (s + t)^2}{2 s t(Q^2-u)}
           \frac{2 Q^2(u-s_2)  + (s + t)^2}{2 s t(Q^2-u)}
%          - \frac{-3 Q^2 + s + t + u}{s t}
          - \frac{s_2-2 Q^2}{s t}
         \Bigg)
\nonumber\\
&
%        - \left[ \frac{s^2 + t^2 + 2 s u + 2 t u + 2 u^2}{s t s_2}
        - \left[ \frac{(s+u)^2 + (t+u)^2}{s t s_2}
             \ln\frac{s_2}{Q^2} \right]_{A+}
        - \ln\frac{\mu_F^2}{s_2} 
             \Bigg(
             - \frac{s t}{(s_2-t)^3}
             + \frac{u+t}{(s_2-t)^2}
\nonumber\\
&
\qquad
             - \frac{3 s^2 - 4 s t + 2 t^2 + 6 s u - 4 t u + 4 u^2}{2st(s_2-t)}
             - \frac{1}{s_2-u} 
%             - \frac{Q^2}{2 u^2} + \frac{Q^4}{2 s u^2} 
%             - \frac{Q^2 t}{2 s u^2} 
             + \frac{Q^2 (u-s_2)}{2 u^2 s}  
\nonumber\\
&
\qquad
%             + \frac{t (s^2 + 2 s t + 2 t^2 + s u - 4 t u + 4 u^2) - 
             + \frac{t ((s+t)^2  +  s u + (t- 4 u)^2) - 
                Q^2 (2 (s^2+t^2) + s t  + 4( s u - t u + u^2))}{st(tu-s_2 Q^2)}     
\nonumber\\
&
\qquad
             + \frac{5}{2 s} - \frac{Q^2}{2 t^2} + \frac{3}{2 t} 
						 - \frac{1}{u} + \frac{Q^2}{s u} 
             - \frac{t}{2 s u}
              \Bigg)
\nonumber\\
&
       + \frac{3 (s^2 + t^2 + 2 s u + 2 t u + 2 u^2)}{4 s t (s_2)_{A+}}
       + \frac{3 s (t - u)^2 (t + u)}{4 t \lambda^4}
\nonumber\\
&
       + \frac{
         -2 t^3 - 4 Q^2 (s - 2 t) (t - u) + 2 t u^2 - 2 s^2 (t + u) + 
                     s (7 t^2 - 10 t u + 3 u^2)}{8 s t \lambda^2}
\nonumber\\
&
       + \frac{t(s^2+su+4(t-u)u)-Q^2(s(t-4u)+4u(t-u))}{2st(tu-s_2Q^2)}
       + \frac{4 s t}{(s_2-t)^3}
\nonumber\\
&
       - \frac{4 u+8 t-3 s}{2 (s_2-t)^2}
       + \frac{-s^2-4st+3t^2+2su-2tu+u^2}{2st(s_2-t)}
       - \frac{s}{t(s_2-u)}
\nonumber\\
&
       - \frac{3}{2 s} - \frac{Q^2}{4 t^2} + \frac{3}{8 t} 
%       + \frac{Q^2}{2 u^2} - \frac{Q^4}{2 s u^2} 
%      + \frac{Q^2 t}{2 s u^2} 
       - \frac{Q^2(u-s_2)}{2 u^2 s}
			 - \frac{1}{u} 
       + \frac{Q^2}{s u} - \frac{t}{s u} \,.
\end{align}

%%%%%%%%%%%%%%%%%%%%%%%%%%%555
\boldmath
\section{Cross section of $q+q\to Z+X$ through $\mathcal{O}(\alpha^2\alpha_s)$}
\label{sec:c}
\unboldmath

In this appendix, we consider $\mathcal{O}(\alpha^2\alpha_s)$ contributions to
$q+q\to Z+X$.
This includes the interference contributions from subprocess (\ref{qq_Zqq})
involving a virtual photon or $Z$ boson.
We have
\begin{equation}
  \frac{s\,d\sigma^{qq}}{dt\,du} = 
    \frac{4\alpha_s\alpha^2 C_F}{N_c s} \left[
            Q_q^2(v_q^2+a_q^2) C_1  
          + (v_q^4+6v_q^2 a_q^2+a_q^4) C_2  \right] \,,
\end{equation}
where
%\begin{eqnarray}
%C_1' &=& 
%%       - A(gs) A(em)^2 [vu2+au2] Qu^2 Kqq 1/s   (
%        + \ln\Big(\frac{(s_2-t)(s_2-u)}{s s_2}\Big)
%            \frac{9(2Q^4+2s^2+2s(t+u)+(t+u)^2-2Q^2(2s+t+u))}{4stu}
%\nonumber\\
%&&
%        + \ln\Big(\frac{s Q^2(s_2-t)^2}{[s_2(2Q^2-u)-Q^2 t]^2}\Big) 
%           \frac{9(2s^2+(t+u-2s_2)(s+Q^2-s_2))}
%                  {8(s+Q^2-s_2)(s_2-t)t}
%\nonumber\\
%&&
%        + \ln\Big(\frac{s Q^2(s_2-u)^2}{[s_2(2Q^2-t)-Q^2 u]^2}\Big) 
%           \frac{9(2s^2+(t+u-2s_2)(s+Q^2-s_2))}
%                  {8(s+Q^2-s_2)(s_2-u)u}
%\nonumber\\
%&&
%        + \frac{9}{4}\ln\Big(\frac{st^2}{Q^2(s_2-t)}\Big) \Bigg(
%            \frac{u-t}{2t(s_2-t)}
%          + \frac{1}{s} + \frac{s-s_2+u}{t^2} + \frac{s_2^2}{stu} 
%          + \frac{t-2s_2}{2su}  + \frac{u-2s_2}{2st}
%\nonumber\\
%&&
%\qquad
%          + \frac{s^2}{t(s_2-t)(s+Q^2-s_2)}
%          \Bigg)
%\nonumber\\
%&&
%        + \frac{9}{4}\ln\Big(\frac{su^2}{Q^2(s_2-u)}\Big) \Bigg(
%            \frac{t-u}{2u(s_2-u)}
%          + \frac{1}{s} + \frac{s-s_2+t}{u^2} + \frac{s_2^2}{stu} 
%          + \frac{u-2s_2}{2st}  + \frac{t-2s_2}{2su}
%\nonumber\\
%&&
%\qquad
%          + \frac{s^2}{u(s_2-u)(s+Q^2-s_2)}
%          \Bigg)
%\nonumber\\
%&&
%         + \frac{9(2 s_2 Q^2 (t^2 + u^2)-tu(t+u)^2)}{4st^2u^2} \,.
%\end{eqnarray}
%ORR
\begin{align}
C_1 &= 
%       - A(gs) A(em)^2 [vu2+au2] Qu^2 Kqq 1/s   (
\frac{s_2^2+(s-Q^2)^2}{tu}
\ln\frac{(s_2-t)(s_2-u)}{s s_2}
      + \frac{2 s_2 Q^2 (t^2 + u^2)-tu(t+u)^2}{t^2u^2} 
\nonumber\\
&
        + s\Bigg\{
%         \ln\Big(\frac{s Q^2(s_2-t)^2}{[s_2(2Q^2-u)-Q^2 t]^2}\Big) 
           \frac{s^2 + (s_2-Q^2)^2}{2(s+Q^2-s_2)(s_2-t)t} L_{\lambda t}
        + \Bigg(
           \frac{s^2 + (s_2-Q^2)^2}{2(s+Q^2-s_2)(s_2-t)t}
\nonumber\\
&
        +   \frac{2s u Q^2 + t(2 s_2 (s-Q^2) + (t+u)^2)}{2s u t^2}
          \Bigg) L_{st} 
       \Bigg\} + s\Bigg\{ u \leftrightarrow t \Bigg\}\,,
\\
C_2 &=
      \frac{s(2 s_2 Q^2 (t^2 + u^2)-tu(t+u)^2)}{t^2u^2(s+Q^2)} 
%\nonumber\\
%&
+ \Bigg\{	 
% -\frac{\ln{(Q^2t-s_2(s+Q^2))^2/(Q^2(s+t)(s_2-t)^2)} }{4( Q^2+s)^2 t u}
-\frac{1}{4( Q^2+s)^2 t u}
\ln\frac{(Q^2t-s_2(s+Q^2))^2}{Q^2(s+t)(s_2-t)^2}
\nonumber \\
&\qquad
   \biggl(Q^2  ( s_2^2+(s_2-t)^2\!-\!u(2 s+u) )+s  (2 s_2(s-Q^2)+(t+u)^2 )\biggr)
\nonumber \\
% &+& \ln\Big(\frac{st^2}{Q^2(s_2-t)^2}\Big)  
 &+ L_{st}
  \biggl[\biggr. 
		\frac{-s^2}{2 (Q^2+s)^2 t}+\frac{s^2}{2 (Q^2+s-s_2) t (s+u)}-\frac{s+t}{4 t (s+u)}-\frac{1}{4 t}
\nonumber \\
 &+\frac{t^2-u^2+2 s_2^2-2s_2 t}{4 (Q^2+s) tu}
   + \frac{s_2 Q^2}{2 t^2 (s+u)}
	\biggl.\biggr]
   + \frac{s^2 +(Q^2- s_2)^2}{4 (Q^2+s-s_2) t(s+u)} \, L_{\lambda t} 
\nonumber \\
&+ \frac{ Q^2 (s_2-u)}{2 u^2 (s+t)} \, \ln\frac{Q^2}{s+t}
  + \frac{1}{4} 
	I_4(1,1,t,1) 
	\biggl[\frac{s^2(Q^2-s-s_2)}{(Q^2+s-s_2) (s+u)} \biggr.
\nonumber \\
 &+   \frac{ (2 s_2^2 + t(Q^2-s_2) ) s}{t (s+u)}
 - \frac{(-2  s_2+t+u)^2}{t} \Bigg]
\nonumber \\
 &+ 
 I_4(1,1,t,-1) 
 \biggl[\biggr.
    -\frac{s^3}{2 (Q^2+s-s_2) (s+u)}
		-\frac{s^2 Q^2}{2 (Q^2+s)^2} 
		+\frac{s(Q^2-u)}{Q^2+s}
\nonumber \\
 &+\frac{s_2 s(u-Q^2)}{2 (Q^2+s) u} 
 -\frac{s (u-t)^2}{4 t ( Q^2+s)} 
 +\frac{s_2 s(u-s_2)}{2 ( Q^2+s) t} 
 +\frac{s^2(2t-s_2)}{4 t(s+u)} 
\nonumber \\
 &-\frac{5 s_2 t}{4 (s+u)}
 +\frac{s_2^2(2t-s_2)}{t (s+u)} 
 +\frac{t^2}{4 (s+u)}
 +\frac{s(2 s_2-t)}{4 (s+u)}
 +\frac{u^2}{2 t}-\frac{5 s_2}{2}
 +\frac{t}{2}+u
\nonumber \\
 &-\frac{9 s_2 u}{4 t} 
 +\frac{3 s_2^2}{t}
 -\frac{s_2^2}{u}
 -\frac{t^2}{4 u}
 +\frac{s_2 t}{u}
 -\frac{5 s_2 s}{4 t}
 +\frac{s_2s}{2 u}
\biggl.\biggr]
 - \frac{ Q^2 s^2}{Q^2+s} I_4(1,2,t,-1) \Bigg\} 
 + \Bigg\{t \leftrightarrow u\Bigg\} \,.
%\nonumber
% + \frac{ s_2 (Q^2-t)}{u^2 (Q^2+s)}+ \frac{2 s_2 u - t (t+u)}{2 ( Q^2+s) t^2}\\
\end{align}
Here, we have again used the phase-space integrals given in
Eq.~(\ref{eq:psintegrals}).

\end{document}